\documentstyle[12pt,aaspp4]{article}
 at 14.40truept
 at 17.28truept
\def \mpc	{{\rm\ Mpc}}
\def \kpc	{{\rm\ kpc}}

\def \kms       {\hbox{ km s$^{-1}$}}

\def \ie	{\hbox{\it i.e.}}
\def \eg	{\hbox{\it e.g.}}
\def \cf	{\hbox{\it cf.}}

\def \etal 	{{\it et al.\ }}

\def \K		{ \hbox{$\,$ K} }

\def \kev	{{\rm\ keV}}
\def \msol	{{\rm M}_\odot}

\def \hinv     	{\hbox{$\, h^{-1}$} }
\def \h50inv	{\hbox{$\, h^{-1}_{50}$} }
\def \ergs	{ \hbox{$\,$ erg s$^{-1}$} }

\def \se	{\!=\!}

\def \sims    {\sim \!}
\def \ssimeq	{\! \simeq \!}

\def \spropto	{\! \propto \!}
\def \grad	{\stackrel{\rightharpoonup}{\nabla}}
\def\aj{{\it A. J.\/}}
\def\\{\hfil\break}
\def\spose#1{\hbox to 0pt{#1\hss}}
\def\lta{\mbox{$^{<}\hspace{-0.24cm}_{\sim}$}}
\def\gta{\,\mbox{$^{>}\hspace{-0.24cm}_{\sim}$} \,}
\def\beq{\begin{equation}}
\def\eeq{\end{equation}}

\def \rhocrit     {\rho_c}
\slugcomment{submitted to the Astrophysical Journal}
\lefthead{Metzler and Evrard}
\righthead{Simulations of Clusters I.}

\begin{document}

\title {Simulations Of Galaxy Clusters With And Without Winds
 I.  The Structure Of Clusters }

\author {Christopher A. Metzler\altaffilmark{1}}
\altaffiltext{1}{metzler@denali.fnal.gov}
\affil {NASA/Fermilab Astrophysics Group, Fermi National Accelerator
Laboratory, Box 500, Batavia, IL 60510-0500 USA}
\medskip
\author {August E. Evrard\altaffilmark{2}}
\altaffiltext{2}{evrard@umich.edu}
\smallskip
\affil {Department of Physics, University of Michigan, Ann Arbor, MI
48109-1120 USA 
and \\
Institut d'Astrophysique, 98bis Blvd. Arago, F75014 Paris, France 
}

\medskip

\begin{abstract}

We use gas dynamic simulations to explore the effects of galactic
winds on the structure of the intracluster medium (ICM) in 
X--ray clusters.  Two ensembles of 18
realizations, spanning a decade in temperature $T$, are evolved with
and without galactic winds in an underlying standard CDM cosmology
with $\Omega \se 1$ and $\Omega_b \se 0.1$.  Galaxies are identified
as peaks in the initial, linear density field, and are assumed to
lose half their initial mass over a Hubble time in winds with
effective temperature $T_{wind} \se 8 \kev$. 

The extra wind energy raises the entropy of the gas above the level
generated by gravitationally induced shocks.  This leads to 
substantially lower central densities in the ensemble with 
winds compared to the ensemble lacking winds.  
The magnitude of this effect increases with decreasing
mass or virial temperature, and results in a trend of shallower gas 
profiles at lower temperatures, consistent with observations.
In contrast, we find the final temperature 
of the gas is relatively unaffected; a similar mass--temperature
relation results with or without winds.  The input wind energy, which
is comparable to the thermal energy in low temperature systems, is
effectively consumed as work to lift the gas in the dark matter
dominated potential.  Radially averaged temperature
profiles of models with winds are slightly steeper than those
without.  The extended nature of the ICM with winds can lead to 
underestimates of the global baryon fraction;  we calibrate the
amplitude of this effect at density contrasts $\delta_c \se 170$ 
and $500$.  These features should be generic to all wind
models. 

The structure of the dark matter density profiles is consistent with
the form proposed by Navarro, Frenk \& White, and we find evidence for 
higher central concentrations in lower mass systems, consistent with
previous, purely N--body studies.  The galaxy distribution in the
ensemble with winds is cooler and more centrally concentrated 
than either the dark matter or gas.  A mild, but persistent, velocity 
bias exists, with ensemble average value $\sigma_{gal} 
\simeq 0.84 \sigma_{DM}$.  

The steep nature of the galaxy spatial 
distribution, combined with ejection of metal enriched material 
over a Hubble time, produces a strong, negative radial gradient 
in metallicity within the ICM.  Core radii remain unresolved, even in
the models with winds.  These features are sensitive to
the assumed wind history of the galaxies.  


\end{abstract}


\newpage

\section{Introduction}

A variety of mature observational techniques are now in
use studying galaxy clusters.  Through optical studies of cluster
galaxies, analyses of weak gravitational lensing distortions of
the background galaxy field, observations of radio sources within
and behind clusters,
and X--ray images and spectra of the intracluster medium (ICM),
we now have a wealth of data to compare to models of clusters
drawn from analytic treatment and numerical simulation.

The paradigm for cluster formation and evolution that has emerged from
such modeling is one in which clusters form through gravitational
collapse of an overdense region (Gunn \& Gott 1972; Bertschinger 1985).
While analytical descriptions typically assume spherical symmetry, 
cluster observations and N--body simulations of hierarchical
clustering from initially Gaussian, random density fields 
show that the collapse process is generally irregular, involving 
mergers of protoclusters flowing along large--scale filaments, along
with accretion of smaller satellite systems and weakly clustered 
material.  

It is commonly held that rich clusters formed at recent epochs.
Nevertheless, since the relaxation timescales for clusters are
significantly less than a Hubble time,
the standard model for describing 
the distribution of matter within clusters is one based on hydrostatic
equilibrium.  Early one--dimensional collapse simulations by 
Perrenod (1978) supported this assumption, later confirmed in 
three--dimensions by Evrard (1990a,b).  The isothermal $\beta$--model
(Cavaliere \& Fusco--Femiano 1976, 1978; Sarazin \& Bahcall, 1977) 
makes further simplifying assumptions of an isothermal ICM temperature 
and spherical symmetry of an assumed, dominant collisionless potential,
now taken to be generated by dark matter.  Each component follows a
density profile of the form 
\beq
\label{eq:betaden}
\rho(r) \,=\,
\rho_{0}\left[1\,+\,\left(\frac{r}{r_c}\right)^2\right]^{-3\alpha/2}
\eeq
where $r_c$ is the core radius within which the density profile
relaxes to a constant, central value $\rho_0$.  In this model, the 
outer profile slopes of the gas and dark matter, measured by 
their respective values of $\alpha$, provide information on the relative 
temperatures of the two components.  The parameter 
\beq
\label{eq:betadef}
\beta\,\equiv\,\frac{\sigma^2}{\left(\frac{kT}{\mu m_p}\right)}.
\eeq
from which the model takes its name, is the ratio of specific energy 
in dark matter, measured by the one-dimensional velocity 
dispersion $\sigma$, to that
in gas, measured by its temperature $T$ and mean molecular weight
$\mu$, with $k$ Boltzmann's constant and $m_p$ the proton mass.  
Since the ICM mass dominates the galaxy mass in rich clusters such 
as Coma (Briel, Henry \& Bohringer 1992; 
White \etal 1993), it is reasonable to assume that 
the ICM plasma originates in primordial gas leftover from galaxy 
formation.  In this case, the gas and galaxies cluster hierarchically 
within the same potential wells, so it is similarly 
reasonable to expect that the specific energies of the two 
components will be nearly equal, $\beta \ssimeq 1$.  
(A refined discussion of this point is provided in the Appendix.) 

However, there is evidence that the history of the intracluster
medium is more complicated.  In particular,
the presence in the ICM of iron
and other elements produced by stars, at abundances near solar,
necessitates significant interaction between galaxies and the
hot intracluster plasma.  Mechanisms for this metal enrichment
process include feedback from a very early stellar population
such as Population III stars
(Carr, Bond \& Arnett 1984), ram pressure stripping by the ICM of the 
interstellar medium from galaxies (Gunn \& Gott 1972; Biermann 1978;
Takeda, Nulsen, \& Fabian 1984; Gaetz, Salpeter, \& Shaviv 1987),
and ejection of hot enriched gas from galaxies via winds (Yahil \&
Ostriker 1973; Larson \& Dinerstein 1975).  How might we discriminate
between these?

First of all, a key question with respect to the dynamics of the ICM plasma
is whether significant energy deposition accompanied the enrichment 
process.  ``Passive'' mechanisms, such as primordial enrichment or
ram pressure stripping, do not add considerable energy to the ICM.
Galactic winds, on the other hand, represent an ``active'' mechanism which 
deposits both energy and metal enriched material into the ICM.  
Meanwhile, there is some evidence implying
cluster gas has a greater specific energy
than cluster galaxies, or $\beta\,<\,1$ (\cf\/ Edge \& Stewart 1991), a 
result consistent with additional, non--gravitational energy input into 
the ICM.  Also,
several studies of the relation between the galaxy velocity 
dispersion and ICM X--ray temperatures in clusters suggest that $\beta$ 
varies with the depth of the potential well (Edge \& Stewart 1991; 
Lubin \& Bahcall 1993; Bird, Mushotzky, \& Metzler 1995; 
Girardi \etal 1995).  To be fair, cluster velocity 
dispersions and X--ray temperatures are difficult to compare in an
unbiased manner, since the quantities are prone to different types of 
systematic errors and are typically not measured within the same region
of a cluster (Metzler 1997).  However, if robust, such a result may be
expected  from wind models.  Since the specific energy
of an individual galactic wind should not depend upon the host cluster
whereas the specific thermal energy supplied by gravitational 
collapse does depend on cluster mass, 
winds should affect more strongly the ICM of clusters with small 
velocity dispersions.  This may introduce a 
dependence of the ratio of specific energies with temperature in 
the manner described above.

Another possible discriminant between enrichment mechanisms lies
in the distribution of metals in the intracluster medium.  However,
it is difficult to infer analytically the type of abundance gradient
expected from each of these three mechanisms.  Simulations of cluster
evolution incorporating enrichment can clarify this, and provide
an expectation to compare to observations of abundance gradients now
becoming available (\eg\/ Tamura \etal 1996; Xu \etal 1997; Ikebe \etal
1997).

We present here results from an ensemble of simulations which include
the effects of galactic winds in a self--consistent, three--dimensional
fashion.  A unique feature of these is the ability to trace the structure
of galaxies and metal--enriched gas in the ICM.  This work expands the
examination of a  single, Coma--like cluster presented in an earlier paper 
(Metzler \& Evrard 1994, hereafter Paper I).  
Since galactic wind models themselves are
uncertain, we take a heuristic approach and employ a simple, and 
in some ways extreme, model for galactic winds in an attempt to 
explore the upper envelope within which realistic models should lie.  
We examine an ensemble of eighteen 
cluster realizations, spanning a factor of 50 in cluster mass, 
drawn from a standard cold dark matter cosmogony.
Each initial realization is evolved twice, with one run
incorporating and the other ignoring galaxies and their ejecta.  
This paper focuses on the three--dimensional structure of the present
epoch population; a subsequent paper will examine the effect of 
feedback on X--ray observations.

In Section 2, we elaborate on the numerical techniques used in this
work, as well as the general properties of the two cluster ensembles
used.  Section 3 provides a look at the structure of the collisionless
components (dark matter and galaxies) in these simulations.  The
structure and metal distribution of the intracluster medium are examined
in Section 4.  A revised model of the ICM, based on the halo model of
Navarro, Frenk \& White (1996, hereafter NFW2), is considered in Section 5.
The relative structures of the various cluster components are compared in
Section 6; also included there are some comments about implications
for estimates of the cluster baryon fraction.  Our results are summarized
in Section 7.

\section{Method}

\subsection{Initial Conditions}

The simulations and their initial conditions use as their basis the
standard biased cold dark matter (CDM) scenario (Blumenthal \etal 1984;
Davis \etal 1985): $\Omega = 1$; baryonic fraction $\Omega_{b}\,=\,0.1$;
Hubble constant $h\,\se\,0.5;$ and power--spectrum normalization
$\sigma_{8}\,=\,0.59$.  These parameters are used throughout this work
when scaling to physical units.  The path--integral formalism of
Bertschinger (1987) is used to generate initial density fields which
are constrained, when smoothed with a Gaussian filter, to have a specified
value at the center of the simulated volume.  For the simulations in this
paper, we filter with a Gaussian of scale $R_{f}\,=\,0.2L \mpc$, where 
$L$ is the length of the periodic volume, corresponding 
to a mass scale of $M_{f}\,=\,\left(2\pi\right)^{3/2}\rhocrit\;R_{f}^{3}
\,=\,5.6 \times 10^{14} \left(L/40\mpc\right)^3\msol$ (Bardeen \etal 1986).
Here $\rhocrit \se 3 {\rm H}_0^2/8\pi G$ is the critical 
density, also the mean background density of the models.  
The perturbation height at the center was constrained to a value
$\delta_{c} \,=\, 2.0$ when filtered on scale $M_{f}$.  For all of
the simulations described in this paper, $32^3\,=\,32768$ particles
are initially placed for each of the dark matter and gas fluids;
the mass of an individual dark matter particle is related to the
mass of a gas particle by $m_{DM}\,=\,9m_{gas}$, reflecting their
fractions of the total density.  The primordial
density field is used to generate a particle distribution at the starting
redshift $z_i\,=\,9$ using the Zel'dovich approximation, as described in
Efstathiou \etal (1985).

Since in generating the constrained initial density field, we filter
on a fixed fraction of the box length, we can simulate clusters spanning
a range in mass simply by varying the box size.  The mass per simulation
particle is proportional to $L^3$, but so is the filter mass scale.  This
causes the number of particles in the final collapsed object to be roughly
comparable in all runs, so the fractional mass resolution in the various
simulations presented here is equivalent.  This avoids any systematics
that might be introduced into correlations between cluster quantities
(X--ray luminosity vs. mass, for example) if the resolution varied
in a systematic way from low--mass to high--mass clusters.

\subsection{Including Galaxies}

The technique used for inserting galaxies in the simulation is described
in detail in Paper I.  We Gaussian--filter the initial conditions
on the approximate scale of bright galaxies ($R_{f}\,=\,0.5\mpc$,
corresponding to
$M_{f}\,=\,\left(2\pi\right)^{3/2}\rhocrit\;R_{f}^{3}
\,=\,1.4\times 10^{11}\msol$) and locate peaks in the initial overdensity
field on that scale above a fiducial threshold of $2.5\sigma$, chosen to
reproduce the observed number density of bright galaxies.  We then return
to the initial particle distribution and replace the gas particles
associated with each peak with a composite ``galaxy particle.''  We assume  
an effective collapse redshift of $z_c\,=\,4.5$, corresponding to a linearly
determined mean interior overdensity at the starting redshift $z_i$ of
\beq
\delta_{gal}\,=\,1.686\frac{1\,+\,z_c}{1\,+\,z_i}\,=\,0.933.
\eeq
The gas particles within this mean interior overdensity are removed, and
the mass of the resulting galaxy particle is set to the number of gas
particles removed.  The initial linear momentum of a galaxy particle is
set by demanding conservation of linear momentum.  

A valid concern with our method and results can be raised over our use
of peaks to simulate galaxies.  The most natural thing to do would be
to allow the gas in the simulations to cool and form galaxies, and then
allow those galaxies to provide the sources for the feedback into the
intracluster medium.  However, such an approach suffers from limitations 
in our ability to accurately model star formation, in both a physical 
and numerical sense.  As we wish to perform
many simulations to ensure adequate statistics when considering issues
of cluster structure and evolution, we must economize computatonal 
resources spent on an individual run, and our approximate peak 
treatment to galaxy formation provides considerable numerical savings. 
The peak model has some physical basis in that 
there is known to be ``crosstalk'' from large to small scales during 
hierarchical clustering from Gaussian initial conditions in the 
non--linear regime which enhances the rate of 
small--scale structure formation for the power spectrum shape 
considered here (White \etal 1987; Juszkiewicz, Bouchet, \& Colombi 1993).  
The model also has some phenomenological success in explaining the 
qualitative shape of galaxy luminosity functions (Evrard 1989) and 
the morphology--density relation in clusters (Evrard, Silk \& Szalay 1990). 

However, since the theory of Gaussian random fields
(Bardeen \etal 1986) tells us that peaks on smaller scales are likely
to be biased towards peaks on larger scales, and since our initial
conditions are constrained to produce a high--peak on cluster scales
at the center of the simulation volume, the initial galaxy distribution
will be more centrally concentrated than the overall mass
distribution.  The thermal history and metal distribution of
the ICM is certainly sensitive to the assumed galaxy formation model. 
To quantify this, several runs were performed with galaxies
placed randomly in the volume, rather than at the locations of overdense
peaks.  By removing the peak correlations induced by the presence of the
cluster, random placement resulted in a substantial reduction in 
the number of bright galaxies within the simulated clusters, even
though the number density in the entire simulated volume was held
fixed.  The effect of feedback was reduced to the point that the 
ejection runs differed little from their non--ejection counterparts,
and so we do not discuss these runs further in this paper.  
High resolution numerical experiments resolving galaxy formation 
within clusters will ultimately settle this question.  The current 
best effort on this issue favors the peaks approach 
over random placement (Frenk \etal 1996).

\subsection{Numerical Algorithm and the Wind Model}

We use the N--body + hydrodynamical algorithm P3MSPH, which combines
the well known particle-particle--particle-mesh ($P^{3}M$) algorithm of
Efstathiou \& Eastwood (1981) with the Smoothed Particle Hydrodynamics
(SPH) formalism of Gingold \& Monaghan (1977).  The combined algorithm
is described in Evrard (1988), and some of the post--simulation analysis
procedures used are described in Evrard (1990b) and Paper I.

The simulation algorithm can follow collisionless dark matter and
collisional baryonic gas; we have modified the simulation algorithm to
also model galaxy particles of varying mass, and to allow the galaxies
included to eject energetic, metal--enriched gas.  The technique used is
described in detail in Paper I.  The galaxy mass fraction
lost through winds is described by a time--dependent rate curve; specific
energy and iron ejection rate curves are also assumed as input.  For
each galaxy, the wind rate curve is integrated until the amount of
mass ejected equals the mass of a simulated gas particle.  Energy and
iron mass fraction are then assigned to that particle by integrating
those curves over the same period.  The process is then repeated
for as long as the ejection rate curve is non--zero.  Thermal energy,
momentum, and iron mass are mixed approximately over the scale of
one SPH smoothing length.  The smoothing process, described in detail in
Paper I, is based on conservation of mass, momentum and energy and a 
scenario in which wind ejecta is rapidly mixed into the surrounding ICM. 
For these simulations, we have assumed a wind model in which 
galaxies eject half their mass at a flat rate from a redshift of 
four to the present, with a wind luminosity for a galaxy with 
$10^{10} \msol$ in baryons of $L_{wind}\,=\,4\times 10^{42} \ergs$, 
and a total energy release of $1.5 \times 10^{60}$ erg.   

\subsection{The Cluster Ensemble}

To study systematic trends, it is necessary to examine an 
ensemble. To this end, we assemble 18 sets of initial conditions,
and evolve them with and without galaxies and winds, for a total of
36 simulations.  Five comoving box lengths are used.  For comoving box
lengths of 20, 25, and 30 Mpc, four sets of initial conditions each are
used, while three each are run at 40 and 60 Mpc.  A summary of general
properties of the runs is shown is Table 1.
As in Paper I, we refer to the ensemble of
runs with galaxies and winds as the EJ, or ejection, ensemble, and
the runs without galaxies as the 2F, or two--fluid, ensemble.
When referring to individual runs in this paper, all run names begin with 
the comoving box length in megaparsecs and end with a suffix  
to differentiate between runs.  We will indicate whether ejection is
included as appropriate.

\section{The Collisionless Components }

\subsection{Clusters Sizes and Characteristic Scales}

In formation via graviational instability, one expects a characteristic 
length to emerge which divides the regions within which material is 
close to hydrostatic equilibrium and exterior to which matter is on 
its first infall or expanding (Gunn \& Gott 1972; 
Rivolo \& Yahil 1984; Bertschinger 1985).  
Because infall occurs on a gravitational timescale 
$t_{grav} \spropto \rho^{-1/2}$, 
one expects this characteristic radius to occur at a fixed value of the 
mean enclosed density.   
Figure~\ref{fig:dm_vrprof} shows the radial velocity profile at
$z\,=\,0.02$ for
the dark matter in four of the two--fluid simulations using 
mean interior density contrast as the abscissa, defined as 
$\delta_c \se \rho(<r)/\rhocrit$.  These four were
chosen because they have qualities worth describing in more detail; 
the remaining clusters have similar structure.  
All show a velocity profile
characteristic of gravitational collapse in an expanding world model.  
Spherical clusters would have a zero velocity surface at a density
contrast of $\sim 5.5$ (Peebles 1980).  As shown by the outer dashed line,
this overdensity does an excellent job of marking the turnaround radius.
In run 20e, the velocity magnitude
in the region of infall is somewhat small, and the infall occurs over
a narrow range of overdensities.  This simulation forms three
small clusters of approximate mass ratios 2:2:1, and the two largest
objects are near each other, causing the infall region in each to be weak
due to interference from the other cluster.  

There is not an obviously sharp transition marking the virialized region.  
Some simulated clusters, such as the 20b and 40a runs shown, 
have a reasonably quiescent region interior to a
region of strong infall.  For these clusters, the rough prediction of the
spherical model --- the inner dashed line at an overdensity of 170 ---
provides a good approximation to the outer boundary of the virialized
region.  Other objects, however, have a complicated velocity structure
within this overdensity.  In particular, the most massive clusters exhibit
infall extending into much larger overdensities.  Massive systems
form later, and these clusters are still experiencing strong infall and are
not relaxed.  The three worst offenders --- runs 40c (shown), 60c, and
60d --- experience strong mergers and asymmetric accretion after a
redshift of 0.5.

Nonetheless, since no other characteristic virial overdensity emerges from
the data, we use the radius with a mean interior overdensity of 170,
hereafter called $r_{170}$, as a fiducial virial radius in the analysis
below.  Cluster properties such as density and temperature will then be
profiled against the scaled radius $x\,=\,r/r_{170}$.  For convenience,
the relation 
between $r_{170}$ and total cluster mass $M_{170}$ is 
\beq
r_{170} = 1.72 \left({ M_{170} \over 10^{15} \hinv \msol} \right)^{1/3} 
\hinv \mpc .
\eeq
If clusters are very nearly self--similar over the range in 
mass probed here, then the choice of another overdensity value 
for the virial scale merely amounts to relabelling the radial coordinate
of our profiles.
Much of the literature follows the example of Navarro, Frenk 
\& White (1995, hereafter NFW1), who employ an overdensity of 200.  
However, Evrard, Metzler \& Navarro (1996, hereafter EMN) demonstrate that  
a density contrast of 500 is a more conservative choice for the 
hydrostatic boundary of clusters, in the sense that the mass weighted 
radial Mach number has smaller variance and an ensemble mean more 
consistent with zero within $r_{500}$ than $r_{200}$.  
For power--law density profiles near $r^{-2}$, $r_{200}$ and 
$r_{170}$ differ by about $8\%$.  

The mass, mean dark matter velocity dispersion, and intracluster 
medium temperature within a radius $r_{170}$ for the members of the
two--fluid ensemble are shown in Table 2.  Although 
the simulations span a factor of 27 in volume, the resulting clusters
span a factor of nearly 50 in mass.  This difference is due to the 
fact that in two of the smallest volume
runs, two clusters of comparable mass form and have not merged by the
end of the simulation.  For the analyses here, the larger of the two
clusters in each simulation was chosen.  
In Table 3, we give information for the ejection ensemble, including 
the global fraction of the initial gas mass remaining in the volume
after insertion of galaxies ($f_{gas}$), 
the number of galaxies in the simulation, the number within $r_{170}$
of the present epoch cluster, and the mean temperature of ICM 
within that radius.  Gas and galaxy fractions within the 
clusters are discussed in \S VI.  The masses and dark matter 
velocity dispersions for the ejection ensemble are very similar 
to their two--fluid counterparts, so we do not quote them here. 

\subsection{Dark Matter Density Profiles}

We now consider the dark matter distribution of the simulated
clusters.  The dark matter structure in the runs with galaxies 
and ejection is nearly identical to their two--fluid counterparts, so
we present results from only the 2F set in this and the 
following section.

Figure~\ref{fig:dm_denprof} shows the dark matter density profiles for
the eighteen clusters in our ensemble, taken at $z\,=\,0.02$.
These profiles were constructed by defining radial bins containing 200
particles each, then measuring the
volume of the bin to arrive at the density.  The shapes of the profiles
look remarkably similar.  In Figure~\ref{fig:dm_denprof_sc}a, the profiles
have been rescaled; we plot the local density contrast 
$\rho /\rhocrit$, versus scaled radius $x\,=\,r/r_{170}$.  
There is some difference
in central overdensity between models, but at larger radii 
(smaller overdensities), this dispersion tightens.  Vertical lines in 
both figures denote the values of the gravitational softening parameter
$\epsilon$ for each individual run at this epoch.  The agreement among
the density profiles of the ensemble reinforces previous 
findings of a characteristic density profile for halos formed 
via hierachical clustering.  The self--similarity displayed in 
this figure confirms the choice of $r_{170}$ as a scale 
radius, although choices of overdensity near this value would work 
equally well.  

Motivated by the self--similar appearance in Figure~\ref{fig:dm_denprof_sc}a,
we  construct a mean density profile for the two--fluid runs  by averaging
the values of the density derived from each individual cluster in radial
bins evenly spaced in $\log\left(x\right)$.  The result, along with 
comparison to various functional forms, is shown in
Figure~\ref{fig:dm_denprof_sc}b.  
Each of these functions has at least two adjustable
parameters --- an amplitude, and either a scale length or an exponent.
However, it is important to note that one parameter is constrained
by the required mean overdensity interior to $r_{170}$.  In fitting
to these functions, only data within $r_{170}$ are used.  

We first consider a fitting function of the form introduced by NFW1 
\beq
\label{eq:nfwden}
\frac{\rho\left(x\right)}{\rhocrit} \,=\,
\Delta \left(\frac{x}{\lambda}\right)^{-1}
\left[1 + \left(\frac{x}{\lambda}\right)\right]^{-2}
\eeq
where, as before, $x\,=\,r / r_{170}$, the scaled radius of
Figure~\ref{fig:dm_denprof_sc}.  This
profile approximates an $r^{-1}$ power law at small radii, and an $r^{-3}$
power law at large radii.  The characteristic scaled radius $\lambda$,
or physical radius $\lambda r_{170}$, is the radius at which the
logarithmic slope of the density profile is $-2$; $\Delta$ is
four times the local overdensity at that radius.  Since we will apply
this functional form to the entire density profile, our integral constraint
requires that
\beq
\Delta\,=\,
\frac
{170}
{3 \lambda^{3}
\left[\ln\left(1+\frac{1}{\lambda}\right) - \frac{1}{1 + \lambda}\right]} .
\eeq
A single member of this class of functions with fixed values 
$\lambda\,=\,0.2$ and $\Delta\,=\,7500$ was introduced by NFW1, 
and shown to model well the inner profiles of their simulated CDM 
clusters.  Subsequent work
(NFW2; Metzler 1995; Cole \& Lacey 1996;
Tormen, Bouchet \& White 1997) generalized 
this profile to allow $\lambda$ to be a free parameter.
When applied to our mean profile, this form provides an excellent fit, with
a best fit $\lambda\,\simeq\,0.154\pm 0.008$
(implying $\Delta\,\simeq\,13600$).  Our normalization looks much
larger than the original NFW1 result but, as explained below, the 
discrepancy is due to differences in the samples employed in the 
studies. 

If we apply this form to two subsets of the ensemble,
one comprised of the six highest--mass runs and one comprised of the
eight lowest--mass runs, we find that the mean profiles are significantly
different,
with the high mass ensemble requiring a higher value of $\lambda$
($0.176\pm 0.010$) than the low mass ensemble ($0.145\pm 0.005$).
A small value for $\lambda$ corresponds to a
steeper inner density profile; low mass CDM halos are more centrally
concentrated than high mass halos.  
The difference in density structure between high and low mass objects
reflects the formation epochs of different objects.  In hierarchical
clustering cosmogonies such as CDM, lower mass objects form earlier, 
when the background density is higher, 
so their mass is expected to be more centrally concentrated.  This
effect is expressed clearly by NFW2, who examine halos spanning four
decades in mass.  It is this mass dependence which 
explains the difference between our best fit parameters and 
the original NFW1 values.  Our fits are, in fact, in good agreement 
with the standard CDM case considered in NFW2 (their Figure~5).  

Contrast the seeming success of this model with the standard
$\beta$--model profile, Equation~\ref{eq:betaden}, which provides
a three--parameter fitting function 
as
\beq
\frac{\rho\left(x\right)}{\rhocrit} \,=\,
\Delta_{0} \left[1 + \left(\frac{x}{x_c}\right)^2\right]^{-3\alpha_{DM} / 2}.
\eeq
This functional form implies a central,
constant density core, characterized by the core radius 
$r_c\,=\,x_c r_{170}$ and central density $\Delta_{0}\rhocrit$. 
Using this
expression, we find a best--fit core radius of $x_c = 0.053$, slightly under
twice the mean softening scale (see Figure~\ref{fig:dm_denprof_sc}).  At
such a radius, the deviation of the softened force from a normal Newtonian
force law is significant, so we cannot claim to resolve such scales in the
mean profile.  Fits of this function to the density profiles of individual
clusters produce resolvable core radii only in systems with recent or
in--progress merger activity. We therefore cannot claim to resolve any
core in our simulated clusters' density profiles, in agreement with
numerous previous studies.  The implied large--radius logarithmic slope
for the mean profile is $-3\alpha_{DM}\,=\,-2.48$.

The simplest description of the density profile is that of a power
law, $\rho(x)  \propto x^{-\alpha}$.  The curvature in the density
profiles evident in Figure~\ref{fig:dm_denprof} implies that a power law
is inappropriate over the entire range of resolved structure, and
formal fits verify its inadequacy.
It is worth noting, however, that while the curvature is clearly
present, it is not extreme.  The local logarithmic slope of the
density profile lies between $-1.5$ and $-2.5$ over the entire
resolved range, lending support to analyses of cluster structure
which assume isothermality.  Considering only radii with a local
overdensity in the range $100 \leq \rho/\rhocrit \leq 3000$, a power
law with $\alpha \se 2.39 \pm 0.08$ provides an excellent fit to
the mean profile.  This result is consistent with the value
$2.33 \pm 0.04$ found  in the $\Omega\,=\,1$, $n\,=\,-1$ model of
Crone, Evrard \& Richstone (1994).  At
smaller radii, the profile is more shallow; a fit between local
density constrasts of $10^{5}$ and 5000 yields $\alpha \se 1.56$.
The spatial and mass resolution of the experiments is not 
sufficient to demonstrate convergence to this, or any other, value 
of the logarithmic slope of the dark matter density as 
$x \rightarrow 0$. 

Parameters extracted from fits to the dark matter profiles
are summarized in Table 4.
There are several important points to summarize.  First, the 
clusters have a characteristic density profile consistent with those 
found in previous studies.  The logarithmic slope of the
profile is typically shallower than -2 at small radii, and
steeper at large radii; the division between these two regions 
occurs between 0.1--0.2 $r_{170}$, depending upon the mass of the 
cluster.  For clusters with emission
weighted X--ray temperatures of $7\kev$ or so, this should
correspond to radii of about 350--700$\kpc$ at the present.  
The degree of central concentration is mass--dependent, with less 
massive clusters being more centrally concentrated.  The outer 
portions of cluster density profiles are well--approximated by 
power--laws and demontrate less sensitivity to mass. 
There is no evidence that CDM clusters have or even approach 
constant density cores.  The behavior of the density profile in the very
central regions of clusters remains uncertain;  recent high resolution
simulations exhibit central profiles steeper than that 
prediced by the NFW form, Equation~(\ref{eq:nfwden}) (Moore \etal 1997).

\subsection{Dark Matter Velocity Dispersion Profiles}

The top half of
Figure~\ref{fig:dm_sigprof_sc} shows the dark matter velocity dispersion
profile for the eighteen members of the two--fluid ensemble.  The
profiles have been rescaled --- the radial
coordinate by $r_{170}$ for each cluster, and the velocity dispersion by
the quantity $\sigma_{170}$, defined as
\beq
\sigma_{170}\,=\,\left(\frac{GM_{170}}{2r_{170}}\right)^{1/2},
\eeq
where $M_{170}$
is the mass within $r_{170}$.  Most of these profiles have a common shape,
rising from the center of the cluster and then falling again towards the
virial radius, but recent merger activity causes deviations from this
profile for some systems.  As noted earlier, the typical dark matter
density profile for the ensemble is shallower than $r^{-2}$ at small radii,
and steeper at larger radii, corresponding to the velocity dispersion profiles
seen.  The radius at which the velocity dispersion is a maximum will lie
somewhat beyond the break radius at which the density profile has a local
logarithmic slope of $-2$.  For the NFW profile, if we assume the velocity
dispersion to vary weakly with radius (true for the simulated clusters),
and that velocity anisotropy is unimportant,
then the location of the velocity dispersion maximum can be calculated to
lie at $x_{max}\,\simeq\,1.16 \lambda$; the mean profile would then predict
the maximum of the velocity dispersion at $x_{max}\,\simeq\,0.18$, very
near where most of the curves in Figure~\ref{fig:dm_sigprof_sc} reach
their maximum.  Deviations
from this prediction for individual curves originates from transients 
associated with mergers and/or the presence of long--lived orbital 
anisotropy in the velocity distribution. 

The bottom half of the figure shows the velocity dispersion
anisotropy parameter, $A\left(r\right)\,=\,1\,-\,\sigma^2_t/\sigma^2_r$,
where $\sigma_r$ and $\sigma_t$ are the dispersions in the radial
and transverse velocities respectively.
The dark matter orbits are mostly radial over much of the profile for
all of the members of the ensemble, reducing the kinetic support somewhat
and steepening the dark matter density profile.  At small radii, the
dispersions converge towards isotropy, although one run (60d) shows
evidence for the irregular state noted in its radial velocity
profile.

\subsection{Galaxy Number Density Profiles}

Representation of galaxies as a separate, collisionless component in 
the ejection ensemble allows us to investigate the kinematics of 
this visible population.  
We fit the distribution of galaxies in the simulated clusters to
a $\beta$--model profile.  Galaxy number density profiles are determined
by constructing Lagrangian radial bins for each simulated cluster, holding
five galaxies each, out to $r_{170}$.  The central two bins of each
profile are excluded from the fit, to minimize the effect of force
softening on the results.  This makes determination of central galaxy number
density and core radius uncertain; but these parameters are of questionable
value, since in real clusters their determination is prone to a variety
of errors, particularly from the choice of cluster center.  We can
estimate the large--radius logarithmic slope of the galaxy number 
density profile, and address the question of 
whether the dark matter is more extended than the galaxy
distribution.  Finally, we consider only cluster profiles which have at 
least eight fitting bins after this exclusion, and thus at least five 
degrees of freedom.  This requires at least 50 galaxies within $r_{170}$.

Figure~\ref{fig:galnumdenprofs} shows the galaxy number density profiles
of the six largest
clusters in the ensemble --- the only six that fit the minimum criteria
above.  Also shown are best--fit $\beta$--model profiles.  
In five of the six cases, the
large--radius slope of the galaxy number density profile $-3 \alpha_{GAL}$ 
is steeper than that of the dark matter; the dark matter is more 
extended than the cluster galaxies.
Although the number statistics here are poor, a comparison of cumulative
masses using the entire ensemble, shown in \S VI, 
clearly demonstrates that the galaxy population is, in the mean, 
more centrally concentrated than the dark matter.  

\subsection{Velocity Bias}

Since our initial placement of galaxies is upon peaks in the density
field, and since such peaks are expected to be spatially biased towards
the peak on large mass scales associated with the cluster itself,
the galaxies are expected to be somewhat more centrally concentrated
than the dark matter.
There are, however, physical mechanisms which can contribute to such
concentration.  Apart from the contribution of
galaxies to the overall cluster potential well, the distribution of
galaxies and dark matter will be affected to some degree by interactions
between the two components (Barnes 1985; Evrard 1987; West \& 
Richstone 1988; Carlberg 1991; Carlberg \& Dubinski 1991; Carlberg 1994).
Given a CDM halo which is initially well--traced by the distribution of
galaxies, dynamical friction will transfer energy from the galaxies to the
dark matter, resulting in a dark matter distribution which is more extended
than would be the case in the absence of galaxies, and a galaxy density
profile which is more centrally concentrated than that of the halo.
A simple timescale argument based on the Chandrasekhar dynamical
friction formula (\cf Binney \& Tremaine 1987) suggests
that, on the periphery of clusters or in the largest clusters,
dynamical friction should be unimportant.  However, in cluster cores and
larger parts of poor clusters, this timescale can be comparable to or
less than a dynamical time.

The effect of such friction on the structure of the dark matter is
small.  If galaxies and
dark matter both have the same initial specific energy, and if each galaxy
loses a fraction $k$ of its initial specific energy through dynamical
friction, the specific energy of the dark matter is boosted by a factor
$\left(1\,+\,k M_{gal}/M_{DM}\right)$.  In rich clusters, galaxies typically
account for perhaps 6\% of the total mass.  If baryons make up $30\%$ of
clusters --- more than suggested by analyses of their mean properties (Evrard
1997) --- then $M_{gal}/M_{DM} \simeq 0.085$.  In this extreme case, even
if galaxies lose as much as 25\% of their specific energy through dynamical
friction, the effect on the dark matter is only 2\%.  Integrating over
the fits to the two--fluid and ejection ensembles' mean dark matter density
profiles confirms that the total and specific energy differences between
the two are less than a couple of percent.  Such an energy gain by the dark
matter is insignificant, and at any rate may be swamped by energy lost
heating the ICM through the varying gravitational potential during collapse
and relaxation.  However, while the effect upon the dark matter should
be weak even if the galaxies lose a large fraction of their kinetic energy,
the actual magnitude of effect upon the galaxies is unclear.  

A possible signal of dynamical friction is the presence of velocity bias
in the cluster, $b_v \!<\! 1$, where 
$b_v\,=\,\sigma_{gal}/\sigma_{DM}$ is the ratio 
of galaxy to dark matter velocity dispersions.  
We examine the evolution of $b_{v}$ for
our simulated clusters, constructing velocity dispersions by averaging
over all the galaxies or dark matter within $r_{170}$, in three dimensions.
For individual clusters, the instantaneous value of $b_{v}$ undergoes
strong fluctuations depending upon the dynamical state of the cluster
at that time.  Even so, the value of the bias parameter is only slightly
above unity (up to $1.05$) for brief periods, and for only a few runs.
We attempt to average out the noise associated
with individual clusters by showing in Figure~\ref{fig:velbias_evol2} the
evolution of $b_{v}$
averaged in each time bin over the entire ensemble of ejection runs, over
the six most massive runs, and the eight least massive runs. In all cases,
velocity bias is clearly present.  
The ensemble--averaged bias parameter, time--averaged
over the period from $z\,=\,0.1$ to
the present, is 0.84.  This value agrees well with an independent 
determination of $b_v$ made by Frenk \etal (1996) using a 
self--consistent treatment for galaxy formation within the cluster.  

The curves imply a mass--dependence to the degree of velocity bias, 
in the sense that more massive clusters are less strongly affected.  
This is consistent with dynamical friction arguments, where the 
braking effect of the dark matter background is 
more efficient in low velocity dispersion environments.            
There is no evidence for a continued decay
in the velocity bias parameter, as would result from dynamical friction.
However, close examination of the $b_v$ evolution curves for individual
clusters shows that they can often be described by a moderate decay,
followed by a jump in velocity dispersion.  The jumps occur when additional
galaxies fall into the virialized volume, boosting the velocity dispersion
with their infall velocities.  With such complicated evolution, it is
unclear whether dynamical friction is actually taking place.  



The observational status of velocity bias in clusters is unclear, primarily
because $\sigma_{DM}$ is, of course, not directly measurable.  If we
define $\beta_{DM}$ as the ratio of specific energies of the dark matter
and gas,
\beq
\label{eq:betadmdef}
\beta_{DM}\,=\,\frac{\sigma_{DM}^{2}}{\left(\frac{kT}{\mu m_{p}}\right)},
\eeq
and if the velocity dispersion for cluster galaxies determined from
observations does not suffer from anisotropies and projection effects (and
these simulations suggest that it would), then $\beta_{spec}$, the
spectroscopic value determined from cluster galaxies, should be related
to $\beta_{DM}$ through the velocity bias parameter,
\beq
\beta_{spec}\,=\,\frac{\sigma_{GAL}^{2}}{\left(\frac{kT}{\mu m_{p}}\right)}
\,=\,b_{v}^{2} \, \beta_{DM}.
\eeq
If the specific kinetic energy in dark matter and thermal energy in cluster
gas are both faithful representations of the cluster potential well depth,
then $\beta_{DM}$ should equal unity.  In this case, the determination of
$\beta_{spec}$ for a cluster would allow determination of its velocity
bias parameter.  This approach was taken by Lubin \& Bahcall (1993), who
examined an ensemble of clusters and calculated the average value of
$\beta_{spec}$ for the ensemble, with the intent of eliminating dependence
on dynamical state through the average.  They  found
$\left<\beta_{spec}\right>\,=\,0.97\pm0.04$, which suggests that
little or no velocity bias is present.  However, this result is subject
to the validity of the assumptions noted above.  Their sample of clusters
demonstrated a correlation between velocity dispersion and temperature,
$\sigma_{GAL}\propto T^{0.6\pm0.1}$.  This result was confirmed by Bird,
Mushotzky \& Metzler (1995), who found $\sigma_{GAL}\propto T^{0.61\pm0.13}$
for a sample of clusters explicitly corrected for the effects of substructure.
Girardi \etal (1996) also obtained a similar result, using an independent
analysis designed to minimize the effects of velocity anisotropies.
While consistent with
$\sigma_{GAL}\propto T^{0.5}$, the power law more strongly suggested by
the data implies that $\beta_{spec}$ is temperature dependent.  This means
that any average value of $\beta_{spec}$
taken from a sample of clusters will depend on the temperature distribution
of the sample,
making its interpretation unclear.  Furthermore, when following
the evolution of an individual cluster, excursions in both $\beta_{DM}$
and $\beta_{spec}$ can occur as a result of mergers.  
Finally, the assumption that
$\beta_{DM}\,=\,1$ implicitly assumes that upon infall, cluster gas
thermalizes very efficiently, and retains little or no energy in
macroscopic motions.  Perfect thermalization is not seen in
simulations; a small fraction of residual kinetic energy in the gas 
is routinely found.  A comparison of 11 gas dynamic codes applied to
a single cluster realization yields a mean and standard error 
$\beta_{DM} \se 1.16 \pm 0.03$ (Frenk \etal 1997).  
Heating of cluster gas through energy input from galaxies drives 
$\beta_{DM}$ to lower values, but with several effects pushing
values larger, a modest velocity bias could still be present.
It should also be noted that the mass--dependence of velocity bias noted
above pushes in the direction of a relation steeper than the virial
prediction $\sigma_{GAL}\propto T^{0.5}$.  In this sense, observational
data on the $\sigma$--$T$ relation
are {\it consistent with} the presence of velocity bias.

\section{The Intracluster Medium}

\subsection{Hydrostatic Equilibrium }

The sound crossing time in cluster gas defines a timescale for the
gas to respond to acoustic disturbances.  For an isothermal,
$\gamma\,=\,5/3$ gas, and with parameters on the
low end of rich clusters, this timecale is 
\beq
t_{cross}\,=\,\frac{r_{170}}{c_s}\,=\,
2.0\left(\frac{r_{170}}{1\mpc}\right)
\left(\frac{T}{10^7K}\right)^{-0.5} {\rm Gyr}
\eeq
For an $\Omega\,=\,1$, $h\,=\,0.5$ cosmogony, a lookback time of $2.0$
Gyr corresponds to a redshift of 0.12.
Since X--ray clusters have been seen to much higher redshift
(\cf\/ Bower \etal 1994; Castander \etal 1994), it seems reasonable to
expect that much of the gas in clusters should be in hydrostatic
equilibrium.  Because the temperature scales with radius as
$T \spropto r_{170}^2$ (EMN), the above timescale is 
independent of cluster size.

Figure~\ref{fig:gas_vrprof} shows a profile of the gas radial Mach number
for two--fluid runs.    The ejection runs are very similar --- the 
main difference being a modest reduction in infall velocities  --- so we 
do not show them here.  The velocities are measured
with respect to the velocity of the center--of--mass of the dark matter
distribution, as in Figure~\ref{fig:dm_vrprof}.  
Again, as in that figure, the multiple--cluster run (20e) displays 
a weak infall region.  The rest
of the curves show infall Mach numbers, averaged over radial shells,
that reach a maximum magnitude of at most 1.2.  Internal
to $r_{170}$, radial motions of the gas are quite weak.  There is, 
however, some modest infall of the gas occurring at $r_{170}$, where 
$\langle v_r/c_s \rangle \simeq 0.2$.  This feature is what prompted 
EMN to suggest $r_{500}$ as a conservative estimate 
of the hydrostatic boundary of clusters.  Within $r_{500}$
there are no significant radial motions of gas in either ensemble.  
Mass weighted mean values of the radial Mach number within $r_{500}$ 
quoted by EMN are $-0.022 \pm 0.022$  and $0.001 \pm 0.016$ 
for the 2F and EJ ensembles, respectively. 
The gas is in hydrostatic balance within $r_{500}$, and very 
near to it within $r_{170}$. 

\subsection{ICM Density Profiles }

The gas density profiles for the individual members of both 
ensembles are shown in scaled fashion in Figure~\ref{fig:gas_denprof_sc_new}.
Like the dark matter, the density profiles of the 2F runs display 
remarkable similarity outside $0.2 r_{170}$.  Although the 
mean profile (bold line in the figure) drops two orders of magnitude 
in this regime, variation about the mean is limited to 
$\lta\, 20\%$.  Dispersion in the central gas densities is much 
higher, and larger by about a factor of 3 than the central variation 
in the dark matter in Figure~\ref{fig:dm_denprof_sc}.
This difference may be physical, 
originating from shocks and sonic disturbances in the 
gas which are absent in the dark matter.  Care must be taken, however, 
since the spatial scales involved are quite close to the minimum 
hydrodynamic smoothing in the experiments.  Higher resolution models 
will be able to address this issue.  For now, we note that 
the cluster with the most diffuse central gas (60c) has a rather 
violent formation history, involving strong merger activity at
low redshift.

The impact of ejection on the gas density structure is dramatic.  
The gas in the EJ runs is much less centrally concentrated than 
that of the 2F ensemble.  
At $0.1 r_{170}$, the average density is depressed by over a factor of 3. 
Power law fits to the mean gas density profiles in the overdensity range
$100\,\leq\,\rho_{gas}/\left(\Omega_b\rhocrit\right)\,\leq\,3000$
(the overdensity range fitted for the dark matter)
produce logarithmic slopes of $-1.75$ (EJ) and $-2.34$ (2F).  

The difference in the mean profile values is driven primarily by 
low temperature clusters with ejection.  
Self--similarity across the mass spectrum probed by the experiments is
strongly broken in the EJ ensemble; there is a systematic change in
ICM structure between low and high mass clusters.  Direct evidence 
for this is shown in Figure~\ref{fig:alpha_T},
where we plot values of $\alpha_{GAS}$ from fits to the standard profile,
Equation~\ref{eq:betaden}, against the mean, mass weighted cluster
temperature $T$
within $r_{170}$ for both ensembles.  Mean values of $\alpha_{GAS}$ 
for the two ensembles are listed in Table~5, along with means for 
clusters hotter and cooler than 4 keV.  The 2F models show no apparent 
trend with temperature, whereas the EJ clusters tend to smaller 
values of $\alpha_{GAS}$, meaning more extended gas distributions, 
at lower $T$.  The trend in $\alpha_{GAS}$ with $T$ exhibited by the
models with galactic winds agrees well with the observed behavior of 
$\beta_{fit}$ with $T$ (Mohr \& Evrard 1997) and 
appears consistent with semi--analytic treatments of galactic wind
input (Cavaliere, Menci \& Tozzi 1997).  

The larger extent of the gas in the EJ clusters  results from the work 
done by the wind energy dumped into these systems.  The trend with
temperature results from the fact that the work done in small clusters
represents a larger fraction of their overall energy budget.  
We next consider the energetics of the ICM. 

\subsection{Energetics and Temperature Profiles }

Figure~\ref{fig:T_M170} shows the mass weighted temperature within $r_{170}$ 
against mass $M_{170}$ within that radius for the two ensembles.  
In contrast to the density 
structure, the striking aspect of the $T-M$ relation is its relative lack 
of sensitivity to galactic feedback.  The 2F ensemble is well fit 
by the solid line $T_{2f}(M) = 4.0 (M/10^{15} \msol)^{2/3} \kev$, while the 
EJ ensemble has a slightly shallower slope ($0.62$) and modest 
($\lta\, 20\%$) upward displacement within the range of total masses
explored.  The dotted line in Figure~\ref{fig:T_M170} 
shows the expectation for 
the ejection run temperature $T_{ej}$ based on assuming 
the wind energy is thermalized and retained within $r_{170}$.  In this
case, energy accounting yields (White 1991) 
\beq
T_{ej}(M) \ = \ T_{2f}(M) \ + \ f_{wind} \, T_{wind}
\label{Tej_T2f}
\eeq
where $T_{2f}(M)$ is the relation from pure infall, $f_{wind}$ is the 
ICM gas fraction injected by winds and $T_{wind}$ is the effective 
wind temperature defined in \S 2.  
The models display no systematic trend of $f_{wind}$ with temperature,
so for the purpose of illustration we use a constant value 
$f_{wind} \se 0.22$.  The expected temperatures exceed the measured 
values over all masses, considerably so at the low mass end.  

The wind energy is not retained as heat in the ICM.  Rather, it is 
used to do work in effectively lifting the gas within the dark matter 
dominated potential.  To substantiate this statement, we calculate an
estimate of the work done on the gas in each run by comparing
the final states of gas in each 2F/EJ realization pair.  Since the 
dark matter which dominates the mass distribution is nearly identical 
in the two runs, we can make an ``instantaneous'' approximation of 
the work done by integrating the change in gravitational potential
energy associated with lifting a gas element from its final radius in
the 2F realization to its final radius in the EJ realization.  
Summing, in a Lagrangian fashion, over radially ordered gas mass 
shells (taking into account the small reduction in gas mass due to
galaxies) produces an estimate of the total work required to perturb
the 2F gas distribution into the EJ configuration for each cluster.  
This estimate of the work required can be compared against a
similarly approximate, ``instantaneous'' estimate of the wind energy 
input input by galaxies within $r_{170}$, 
$E_{inp} \se 3/2 M_{gal} kT_{wind}$, where $M_{gal}$ is the galaxy
mass within $r_{170}$.

Figure~\ref{fig:work_Mvir2f_paper}a shows the result of this exercise.
The agreement between 
these two ``instantaneous'' measures is quite good for most of the
clusters.  There is a systematic trend apparent; the slope of the
points is evidently steeper than unity.  We do not fully understand the
cause of this steepening, but speculate that it may be connected to
the difference in formation histories discussed in \S 3.  Given 
the approximate nature of this calculation --- assuming a static
potential well when, in reality, heating of the gas occurred within 
the evolving potential over nearly a Hubble time --- it is perhaps 
surprising that the agreement for most clusters is as good as it is.  
In poor clusters, the estimated work done exceeds the total thermal
energy of the cluster gas affected, as shown in
Figure~\ref{fig:work_Mvir2f_paper}b.  
For systems with total mass $M_{170} \,\lta\, 3 \times 10^{14} \msol$ 
($T \,\lta\, 4 \kev$), the work estimate is comparable
to, and in a few cases exceeds, the thermal energy of 
the gas.  Given the magnitude of the input energy, it is remarkable 
how little net thermal heating occurs, as displayed in Figure~10.  

How much is the internal temperature structure affected by winds?  
Figure~\ref{fig:gas_Tprof_sc_new} shows the temperature profiles for the
members of both ensembles at $z\,=\,0.02$, scaled by 
a fiducial virial temperature,
\beq
T_{170}\,=\,\frac{GM_{170}}{2r_{170}}
\,=\,7.57\times 10^6 \left(\frac{r_{170}}{1\mpc}\right)^2
\left(1\,+\,z\right)^3\,K
\eeq
There are clear structural similarities in the temperature profiles of the 
two samples; both display approximately isothermal behavior 
within half the virial radius followed by a drop to about
half the central value at $r_{170}$.   There is a fair amount of
dispersion at small radii and evidence for a modest central temperature
inversion in some of the 2F systems.  Such a temperature inversion may
be expected from the shape of the dark matter velocity dispersion
profiles; the density profile is shallower than $r^{-2}$ at small
radii.  

There are some structural differences as well.   
The EJ profiles are slightly ($\sims 25\%$) hotter in the 
central regions than the 2F models.  This offsets the lowered density 
in these models and maintains hydrostatic balance.  The central 
pressures in the EJ runs are smaller by factors of $2-3$ than
their 2F counterparts, but the thermal pressure gradient supporting 
the gas ($- \grad \! P/\rho$) is similar in the models.  
One observable consequence of ejection is a slightly steeper
temperature profile, a feature for which there appears to be some
empirical support from ASCA observations of clusters (Markevitch 1997).  

Note that constructing temperature profiles by averaging over spherical
shells can mask significant structure in the temperature distribution.
For example, EMN present in their Figure~1
a highly irregular projected temperature map for a simulated cluster
which, when averaged over three--dimensional shells or two--dimensional
annuli, appears to have an isothermal temperature profile.

Finally, we turn to the energetics of the ICM with respect to the dark
matter.  Figure~\ref{fig:betatruehists} shows the distribution 
of values of $\beta_{DM}$, defined by Equation~(\ref{eq:betadmdef}), 
measured within $r_{170}$ for the two ensembles.  
We also construct a second set of values for $\beta_{DM}$ for
each ensemble, where the temperature is replaced by a total 
specific energy,
\beq
\label{eq:teff}
T\,\rightarrow\,T\,+\,\frac{\mu m_p \sigma_{gas}^2}{k},
\eeq
to take into account the bulk motions (including rotation) and 
residual velocity dispersion of the gas caused by 
mergers and infall.  The figure shows average values for $\beta_{DM}$
for each data set.  
Because most of the wind energy goes into redistributing the gas, 
the shift in the mean $\beta$ values between the EJ and 2F models 
is only $10\%$, comparable to the effect of including gas kinetic energy.
Note also that if $\beta_{DM}$ correlates with cluster temperature,
as may be true for the observed $\beta_{spec}$, then our average
values of $\beta_{DM}$ depend on the cluster sample used.  The
important point here is not the exact values of these averages,
but the relationships between them.

\medskip
\subsection{Iron Abundances and Abundance Gradients}

The gas ejected by our ``galaxies'' is metal enriched.  The distribution
of metals in the ICM of our simulated clusters can thus be examined and
compared with observations.  Although predictions for the metal
distribution are a feature unique to the ejection models, the
predictions themselves are not unique, but depend on the choice of
ejection history, as shown in Paper I.  

We noted above that the gas distribution is more
extended than the dark matter distribution in the ejection runs, with
$-1.75$ as the best--fit power law slope at overdensities
$100\leq \rho_{gas}/\left(\Omega_{b}\rhocrit\right)\leq 3000$.
Meanwhile, in  Figure~\ref{fig:galnumdenprofs}, we saw that
the values of $\alpha_{GAL}$ from fits to the galaxy number
density profiles for individual runs were typically significantly
higher, with only a small overlap in range of values.  The gas is thus
considerably more extended than the galaxy distribution.  
In Paper I, we showed how
this can lead to an abundance gradient; the ejected, enriched gas
traces the galaxies, and thus has a gradient with respect to primordial
gas.  Figure~\ref{fig:Feprof} shows that such a gradient is generally 
seen in the runs from about $x\,=\,0.15$ out to the virial radius.  

The flat metallicity profiles at smaller radii are not induced by
any gravitational or hydrodynamic force resolution effects.  Since
resolution limits flatten the gas density profile at a larger radius
than the galaxies, we would expect this to actually steepen the
gradients near the center.  Instead, the flattening is 
due to the mixing of 
metals at the time of gas particle ejection, which effectively 
acts as a diffusive term.  Many gas
particle ejections take place at radii below $0.1$ $r_{170}$, and the
core gas consequently undergoes many such mixing events.  This was
noted in Paper I (Figure 5b), where we showed that a model with
no mixing evidenced a steeper abundance profile at small radius.  An
ejection model different from the one used here, in which the metal
enrichment took place at earlier times, was also shown in Paper I 
to exhibit a flatter central abundance profile; this occurs because the
difference in profiles between galaxies and gas was not yet large
when the enrichment took place.

Data on abundance gradients from real clusters are only now becoming
available through observations with the ASCA satellite.
These data suggest that clusters sometimes have gradients and
sometimes do not and that poorer systems are more likely
to show evidence for an abundance gradient (Mushotzky 1994; 
Xu \etal 1997).  
Splitting our sample into high--mass and a low--mass
subsets reveals no substantial difference in their gradients, as 
shown by the dotted and dashed lines in Figure~\ref{fig:Feprof}.  

Recall that the abundances in Figure~\ref{fig:Feprof} are scaled by the wind
abundance.  If the wind abundance is constant in time, its absolute value
will not affect the shape of the abundance gradient.  Similarly, a change
in the total quantity of mass blown out of galaxies will not change the
shape of the gradients, as long as the ejection rate remains flat.  What
{\em will} affect the gradient shape is if the metallicity of the wind or
the ejection rate from galaxies vary with time.  If the observations
continue to support weak abundance gradients in rich clusters, 
then the discrepancy between these simulations and the
observations implies that an ejection model in which the
enrichment and heating took place predominantly at early times 
would be more appropriate.  

By late times, the gradient between galaxies and gas that we see
in our simulations should be present in both low--mass and high--mass
clusters.  Late--time enrichment should thus result in abundance
gradients for both mass ranges.  However, since low--mass systems
form earlier in hierarchical structure models, early--ejection
models would enrich the ICM when a galaxy/gas gradient is in place
for low--mass clusters, but not for the high--mass clusters we see
at low redshift.  This would explain such an observational distinction.
There is some evidence from observations of Abell 370 ($z\,=\,0.37$)
that enrichment takes place primarily at early times (Bautz \etal 1994);
the case for this is strengthened by the ASCA detection of significant
quantities of neon, silicon, and other heavy elements in the ICM
(Mushotzky \etal 1996) in quantities which imply injection of
Type II supernova enriched gas rather than Type I.  
This, in turn, implies that the feedback took place early in the lifetime
of cluster galaxies.  We await additional ASCA and upcoming AXAF data,
as well as simulations with ejection based on well motivated star 
formation histories, to clarify this issue.

\section{A Model for ICM Structure and the Core Radius Question }

The traditional formalism used in describing the cluster gas distribution
is the hydrostatic isothermal $\beta$--model, reviewed in
the Appendix.  However, it
is somewhat disconcerting that the fundamental assumptions
upon which the $\beta$--model is based are probably wrong.  In particular,
despite the presence of cores in X--ray images, there is no evidence
for the presence of a core in cluster potentials examined through
gravitational lensing observations, and simulations of clusters in both
CDM and scale--free cosmologies show no core in the underlying dark
matter distribution.  The success
of the model in fitting X--ray surface brightness profiles may be due
to having three free parameters in the fitting function, when there are
three basic features to fit in the data --- an amplitude, a scale
length to define curvature, and a large radius slope.  In addition,
the choice of cluster center is often adjusted to provide the best fit,
introducing additional degrees of freedom.  
The success of the $\beta$--model
profile function in modelling the gas distribution need not require that
its underlying assumptions about the form of the potential be valid.

A second model can be constructed from what we learned in the previous
section on the dark matter density.  The one--parameter form introduced
by NFW provides a good fit to the dark matter mean density profile over
all resolved radii; integrating to find the dark mass within a given
radius, we have
\beq
\label{eq:nfwmass}
M_{DM}\left(< x\right)\,=\,
4\pi\rhocrit r_{170}^3\Delta\lambda^3
\left[\ln\left(1\,+\,\frac{x}{\lambda}\right)\,-\,
\left(\frac{x}{\lambda}\right)\left(1\,+\,\frac{x}{\lambda}\right)^{-1}\right]
\eeq
where recall that $x\,=\,r / r_{170}$, the radius of interest as a
fraction of our fiducial virial radius.  This gives the mass within
a scaled radius $x$, and thus within a physical radius $r\,=\,x r_{170}$.
If we make the assumption that the dark mass essentially determines
the cluster potential (approximately true in our simulations where the
baryon fraction of the simulation volume is 10\%), we can rewrite
this expression as
\begin{eqnarray}
M\left(<x\right)\,=\,M_{170}
\frac
{\left[\ln\left(1\,+\,\frac{x}{\lambda}\right)\,-\,
\left(\frac{x}{\lambda}\right)\left(1\,+\,\frac{x}{\lambda}\right)^{-1}\right]}
{\left[\ln\left(1\,+\,\frac{1}{\lambda}\right)\,-\,
\left(1\,+\,\lambda\right)^{-1}\right]},
\end{eqnarray}
Here $M_{170}\,=\,\frac{4\pi}{3}r_{170}^3\times170\rhocrit$ is the
mass within $r_{170}$.  For an isothermal gas in hydrostatic equilibrium
in the potential defined by this mass profile, we then have
\beq
\frac{1}{\rho_{g}}\frac{d\rho_{g}}{d\left(\frac{x}{\lambda}\right)}
\,=\,-K\left(\frac{x}{\lambda}\right)^{-2}
\left[
\ln\left(1\,+\,\frac{x}{\lambda}\right)
\,-\,
\left(\frac{x}{\lambda}\right)\left(1\,+\,\frac{x}{\lambda}\right)^{-1}\right]
\eeq
where $\rho_g$ is the local gas density, and the constant $K$ is
\begin{eqnarray}
K & = & 4\pi G \rhocrit r_{170}^2\left(\frac{kT}{\mu m_{p}}\right)^{-1}
\Delta\lambda^2 \\
& = & \frac{G}{\lambda r_{170}}\left(\frac{kT}{\mu m_p}\right)^{-1}
\frac{M_{170}}{\left[\ln\left(1\,+\,\frac{1}{\lambda}\right)\,-\,
\left(1\,+\,\lambda\right)^{-1}\right]} \\
\label{eq:Kbeta}
& \simeq & \frac{2 \beta_{DM}}{\lambda\left[\ln\left(1\,+\,\frac{1}{\lambda}\right)\,-\,
\left(1\,+\,\lambda\right)^{-1}\right]},
\end{eqnarray}
where here we have written $\sigma_{DM}^{2}\,\simeq\,GM_{170}/2r_{170}$.
In the absence of any
significant post--infall heating or cooling of the intracluster gas,
we expect the gas temperature to reflect the potential well depth, and
thus $\beta_{DM}\,=\,1$.  Note that even without such additional physics,
we do not expect $\beta$ to equal unity locally,
because we have assumed the gas to be isothermal, while we have shown
earlier that this potential generates velocity dispersion profiles which
are not strictly isothermal.  However, K depends only on the global
value; we consider the impact of such local deviations upon the global
value to be small, and make this assumption in the interests of simplicity.

Integrating over a range in radii from $x_{1}$ to $x$ gives
\beq
\ln\left(\frac{\rho_g\left(x\right)}{\rho_g\left(x_{1}\right)}\right)
\,=\,
K\left(\frac{x}{\lambda}\right)^{-1}\left[
\ln\left(1\,+\,\frac{x}{\lambda}\right)
\,-\,
\left(\frac{x}{x_1}\right)\ln\left(1\,+\,\frac{x_1}{\lambda}\right)
\right].
\eeq
This can be rearranged to give
\beq
\label{eq:nfwgasden}
\rho_g\left(x\right)\,=\,
\rho_g\left(0\right)\,
\exp\left\{
K\left[
\left(\frac{x}{\lambda}\right)^{-1}\ln\left(1\,+\,\frac{x}{\lambda}\right)\,-\,1
\right]
\right\}
\eeq
Clusters in this two--component model 
are essentially a two--parameter family, determined by 
$\lambda$ and $T$.  The latter sets $K$ through Equation~(\ref{eq:Kbeta}). 
Given $K$, the normalization $\rho_g\left(0\right)$ is determined by the
total gas content of the cluster.

In a cosmological setting, the shape parameters $\lambda$ and $T$ are not 
formally independent, as they are both related to the cluster mass.   
This implies that clusters are essentially a one--parameter family, with 
internal structure determined by their mass.  
However, one must be cautious not to oversimply the picture.  
First, as noted before, there is considerable scatter in values of
$\lambda$ at fixed mass --- Tormen, Bouchet \& White (1997) quote a
factor of 2 at the $2\sigma$ level --- which presumably arises from
particular differences in clusters' dynamical histories.  Also, scatter
in the relation between temperature and mass of $10 -20 \%$ arise from
mergers (EMN), and other heating and cooling mechanisms 
may increase this scatter.  Finally, the gas is not exactly
isothermal, as assumed in the model.  Nevertheless, we present this
model, if only as an alternative to the standard $\beta$--model.  
A clear advantage over the latter is that the potential assumed  
in deriving Equation~\ref{eq:nfwgasden} is motivated by direct
simulation of hierarchical clustering.  

Like the $\beta$--model profile, Equation~\ref{eq:nfwgasden}
has zero logarithmic derivative at the origin.  At the
$\beta$--model profile's core radius, the density has dropped to
$\rho_{GAS}\left(0\right)/(2^{3\alpha_{GAS}/2})$, or slightly less than half
its central value for observed values of $\beta$.  If we define a
core radius $x_{1/2}$ for our density profile as the
radius at which the gas density drops to half its central value, we
have
\beq
\label{eq:nfwgascore}
\left(\frac{x_{1/2}}{\lambda}\right)^{-1}
\ln\left(1\,+\,\frac{x_{1/2}}{\lambda}\right)
\,=\,1\,-\,\frac{\ln 2}{K}.
\eeq
This is a transcendental equation for $x_{1/2}$ as a function of $\lambda$.
For the value of $\lambda\,=\,0.154$ from the fit to the mean dark matter
profile, and for the an approximate $\beta_{DM}\,=\,1.17$ for the
two--fluid ensemble, we have $K\,=\,13.2$, which in turn implies
$x_{1/2}\,=\,0.11\lambda\,=\,0.017$.  For a cluster with a $3\mpc$ virial
radius, this corresponds to a core radius for the gas of $51\kpc$, too
low to compare
favorably with observations.  Alternately, using the definition of $K$
earlier, a core radius $x_{1/2}\,\simeq\,0.1$ (corresponding to gas core
radii 100--400 \kpc) implies $\lambda\,\simeq\,0.6$, much larger than is
seen in the simulations.  

Figure~\ref{fig:gasdenprofs5} shows the ICM mean density profile
for the ensemble.  The large dispersion in central values is
reflected in the large error bars for the central points.  The
vertical lines denote the values of the central SPH smoothing length
for the members of the ensemble.  Also
shown is a fit to Equation~\ref{eq:nfwgasden}.  The success of the
fitting function is quite striking, but the best--fit
$\lambda\,=\,0.26$ is inconsistent with the best--fit
$\lambda\,=\,0.154$ extracted from the mean dark matter profile.  For
contrast, the profile for $\lambda\,=\,0.154$ is shown as a dashed line;
the normalization for this curve is set by the average baryon fraction
within $x \se 1$ for the ensemble.  The core
radius $x_{1/2}$ produced by the best--fit profile
with $\lambda\,=\,0.26$ is $0.034$, still too small compared to
observations, although better than the curve inferred from the dark matter
distribution.

In Section III, we established that the NFW functional form provided a good
description of the dark matter profile.  Earlier in this section, we
showed that the gas in the simulations can be thought of as isothermal
and in hydrostatic equilibrium.  And yet here, the gas density profile
that theoretically should be a simple consequence of these facts is
found to conflict with the gas density profile in the simulations.  Why
does the model fall short with the gas, when it succeeds with the dark
matter and when the assumptions about the gas upon which it is based
seem valid?  To answer this, we first note that the shape of
Equation~(\ref{eq:nfwgasden}) depends both on $\lambda$ and $K$, which is 
determined by the temperature $T$ through the hydrostatic equation.  
Figure~\ref{fig:betatruehists} indicates that post--infall thermalization
of cluster gas is
incomplete and that modest residual bulk motions exist in the gas.  
In this case, the gas temperature alone slightly underestimates the 
degree of pressure support.  Incorporating these motions into an
effective temperature, Equation~(\ref{eq:teff}), leads to a decrease 
in $K$ and a gas distribution which is more extended than
that derived from the thermal temperature alone.  From
Figure~\ref{fig:betatruehists}, using
$\beta_{DM}\,=\,1.03$ instead of $\beta_{DM}\,=\,1.17$ results in 
the dash--dotted curve shown in
Figure~\ref{fig:gasdenprofs5}.  (Again, the normalization
comes from forcing the baryon fraction to the average value for the
ensemble).
At large radius, the agreement with data from the simulations is quite good.
Since bulk motions {\it should} be included as a
source of support, this curve is the one of interest.  Our question has
therefore changed:  why does the theoretical
prediction based on the potential and the complete energy budget of the
ICM overestimate the central gas density when constrained to accurately
describe the gas density at large radius?  Why is the gas more extended
in the simulations than is predicted?

To address this, we fit the gas density profiles of individual two--fluid
runs to the $\beta$--profile and extracted core radii.  The top row of
Figure~\ref{fig:gasden_corehists}
shows the distribution of core radii obtained,
the relation between the resulting values of $r_c$ and the corresponding
clusters' virial radii, and between $r_c$ and the numerical parameters
$\epsilon$ (the gravitational softening) and $h_{cen}$, the SPH smoothing
length at the cluster center.  Of principal importance is the latter.
The cluster core radius resulting from a fit to the gas density is
typically three times the central value of the SPH smoothing length.
This is the approximate effective width of the smoothing kernel used
in the simulations; when volume--weighted, the kernel drops to 10\%
of its maximum value at 2.2$h$; it drops to 1\% at 2.8$h$ and 0.1\%
at 3.2$h$.  We argue therefore that the theoretical model, when the entire
ICM energy budget is considered, provides a good description of the
large--radius behavior of the gas density in the simulations.  Its failure
to compare well to the simulations at smaller radii largely 
reflects the fact that the gas density cores seen in the simulations
are principally numerical in origin.  Since the hydrodynamical resolution
in these simulations is comparable to most other studies, and since no
physics beyond shock heating exists in such simulations to raise the
adiabat of cluster core gas, we suggest that gas cores seen in these other
studies are also numerical in origin.  Support for this position is found
in recent numerical studies 
by Anninos \& Norman (1996), which were not able to converge to a 
well--defined gas density core as resolution improved.  It remains 
possible that physical effects contribute to the difference between 
the model and simulated core profiles.  In particular, deviations 
from isothermality and small amounts of core rotational support are 
present in the simulated clusters but not in the analytic model. 

In the right--hand--side of Figure~\ref{fig:gasdenprofs5}, we compare
the NFW gas density
profile to the mean profile for the ejection ensemble.  Once again,
the best--fit does an excellent job of reproducing
the mean profile, but with a scale length of $\lambda\,=\,0.329$, much
larger than the scale in the potential.  The dashed and dash--dotted
lines are as before; thus, the dash--dotted
line contains the bulk motions of the gas indicated in
Figure~\ref{fig:betatruehists}, and is the true prediction of the
theoretical model.  Once again, the core in the real density profile is
more pronounced than that of the model, although not as severely
as for the two--fluid runs.  The ICM in the ejection runs has
experienced post--infall heating, slightly raising ICM gas 
temperatures over pure infall runs, and resulting
in $\beta_{DM}$ decreased typically by 8\%.  This translates into
a more extended gas distribution.  However, as is seen in the figure,
the heating provided by our
ejection model is not sufficient
alone to account for the ICM cores we see in the simulation runs;
numerical effects must still be important.  This is illustrated
by the vertical lines on the figure which mark the value of $h_{cen}$
for the various members of the ensemble; again, $3h_{cen}$ approximately
marks the range of interest, where deviation from the theoretical
prediction begins
to be important.  The bottom row of Figure~\ref{fig:gasden_corehists} shows
that the cores still have a lower limit of $3 h_{cen}$, but now extend
to larger radii as well.  Numerics still dominate the core radii in
the ejection ensemble, but the entropy introduced by winds is 
beginning to play a role.
We conclude that the model employed here is not sufficient
to generate the depressed central densities and cluster
core radii necessary to compare to real systems.

We could have predicted this result.  Equation~\ref{eq:nfwgascore} gives
the dependence of the core radius in this model on the parameter $K$.
For $\lambda\,=\,0.154$, correcting $K$ by a factor of $1.03/1.17\,=\,0.88$
(the ratio of values of $\beta$, or of temperatures) merely
raises the size of the core radius by 10\%.  While additional sources
of heating raise the temperature, thus lowering $K$ and thus increasing
$x_{1/2}$, the increase we need is extremely large.  To raise $x_{1/2}$
to values comparable to real clusters (\ie\ $x_{1/2}\,=\,0.1$), the
temperature must be boosted by a factor of
three or four; such a model is physically implausible and observationally
unjustified.

We remain without an explanation of the gas cores of real clusters.  The
cores predicted by the analytic model, based upon the application of the
NFW profile to our simulated mass distributions, appear too small to
explain those observed in real clusters.  One possibility is
that the mass density profile for real clusters is typically shallower
than indicated by the NFW profile here, in the sense that it approaches
$r^{-2}$ at a larger radius than occurs in these simulations.  This would
be the case if the parameter $\lambda$, the scale radius of the NFW model
in units of $r_{170}$, were significantly larger than is seen here, as
may be the case for CDM models with lower values of $\Omega_{o}$
(Navarro, Frenk \& White 1997).
Another possibility, however, is
that additional physics, such as magnetic fields, provide an important
source of support at small radius.  This view is bolstered by measurements
of Faraday rotation in X--ray clusters (Taylor, Barton \& Ge 1994;
Ge \& Owen 1994) and comparisons of X--ray and lensing mass
measurements for cluster core regions (Loeb \& Mao 1994; Miralda-Escude
\& Babul 1995).  Finally, winds with energy output concentrated at
early times may raise the central entropy to a higher level than
that seen in the models used here, and this may be sufficient to
generate core radii of the requisite scale.  Accurate modelling of 
these effects awaits future simulations.

\section{Relative Structure and the Baryon Fraction}

We summarize in Table 6 what we have learned about
the relative extents of gas and dark matter by showing the results
of power law fits to the outer slope of the mean ensemble density profile
for each fluid in the two ensembles.

For the dark matter, we do not quote separate values for the two
ensembles as their outer slopes are not significantly different.
Small number
statistics do not allow a comparable fit for the galaxies; the range of
values quoted in Table 6 come from Figure 5.  We can show their relative
extent more clearly by displaying in Figure~\ref{fig:relextents_new}
the {\it cumulative}
density measure --- that is, the fraction of the virial mass in each 
component found within a given radius.     
This figure illustrates that the gas with or without 
feedback is more extended than the dark matter, and that 
the difference is significantly enhanced by the energy
input from galactic winds.  At small radii,
the dark matter in the ejection ensemble is typically slightly more
extended than the corresponding dark matter in the two--fluid runs;
however, the difference is quite small.
Finally, the galaxy profile is more centrally
concentrated than any of the other components.  The half--mass radii
for each of the fluids displays this hierarchy.  

As a consequence of the extended nature of the gas distribution, 
the mean, enclosed baryon fraction is reduced relative to 
the global value $\Omega_b / \Omega_0$ at radii interior to the virial
radius.  In the EJ ensemble, the ICM mass fraction is further reduced
by the baryons incorporated into galaxies.  The amplitude of this
reduction within radii encompassing density contrasts 
$\delta_c \se 170$ and 500 is shown in
Figure~\ref{fig:fb_T_paper}, where 
we plot the normalized, local baryon fraction of 
each component, defined by  
$\Upsilon_X \se  f_X (\Omega_b/\Omega_0)^{-1}$ with 
$f_X \se M_X(\delta_c)/M_{tot}(\delta_c)$ the mass fraction of
component $X$ (gas, galaxies, total baryons) within the 
radius encompassing density contrast $\delta_c$.  

The ensemble without ejection displays modest diminutions, 
$\langle \Upsilon \rangle \se 0.94$ within $r_{170}$ and $0.91$ 
within $r_{500}$, with no apparent trend
with cluster size.  The slight differences in formation
history between low and high mass systems, which produces the modest
structural differences discussed in \S III, is not apparent in the
behavior of gas mass fraction with temperature.  The ensemble with ejection
exhibits markedly different behavior.  Although the galaxy mass 
fraction is insensitive to cluster size, the ICM mass 
fraction is noticeably reduced in lower temperature systems, 
particularly those with $T \,\lta\, 4 \kev$.  The magnitude of the effect 
depends on the region under consideration.  At $\delta_c \se 170$, there
is a $30\%$ fractional drop (from $\Upsilon_{gas} \se 0.75$ to $0.55$)
between 10 and 1 keV, whereas the effect is nearly a factor two 
drop at $\delta_c \se 500$.  The larger effect at higher densities or
smaller radii reflects the difference in ICM structure between low and
high $T$ objects displayed in Figure~\ref{fig:alpha_T}; lower temperature
clusters in the ensemble
have less centrally concentrated gas distributions.  At smaller
radii or higher density contrasts, the disparity in gas fractions
between low and high $T$ clusters increases.  

For high temperature clusters ($T \,\gta\, 4 \kev$), the total baryon 
fraction within $r_{170}$ is nearly unaffected by galactic winds.  
There are,
however, modest structural differences in the gas distribution within 
this radius, as indicated from the reduction in $\alpha_{gas}$ from
values $\sims 0.89$ to $\sims 0.79$ (Table~5).  
The work associated with the 
wind energy injected into rich clusters 
is thus used to redistribute gas within $r_{170}$, but causes
little or
no ``spillover'' at this radius.  This is not the case for the low
temperature clusters, where the considerable wind energy results in 
a more dramatic redistribution of the gas extending beyond $r_{170}$.
Still, the diminution of the total baryon fraction is not
catastrophic, even at low temperatures.  A crude fit to the total
baryon fraction $\Upsilon(T)$ with $T$ (in keV) for the EJ 
ensemble yields 
\beq
\label{eq:fb_T_fit}
\Upsilon(T) \ \simeq 1 - A T^{-2/3}
\eeq
where $A \ssimeq 0.3$ at $\delta_c \se 170$ and $A \ssimeq 0.4$ at 
$\delta_c \se 500$.  These are shown as solid lines in
Figure~\ref{fig:fb_T_paper}. 

We stress that the exact form and magnitude of $\Upsilon(T)$ is 
likely to be sensitive to the specific wind model employed.  
One must look for empirical and/or additional theoretical support 
to justify a particular model.  Our particular implementation  
is successful at reproducing the slope of the observed
X--ray size--temperature relation (Mohr \& Evrard 1997), and 
this may indicate that our particular wind implementation is well 
calibrated, in terms of its energetic or entropic effects.  
This provides some reason to be optimistic that the predicted form 
for $\Upsilon(T)$ in Equation~\ref{eq:fb_T_fit}
may apply to real clusters.   Future efforts employing
more realistic wind ejection histories are needed to clarify this
issue.

\section{Summary}

We investigate the structure of clusters in a CDM universe using 
multi--component, numerical simulations spanning a factor of 50 in 
cluster mass.  Dark matter, intracluster gas 
and galaxies are included in a set of models designed to explore the 
role of feedback of mass and energy into the ICM.  Two principal 
assumptions made for the galaxy population are: (i) galaxies form at 
the locations of peaks in the initial density field and (ii) galaxies 
lose half of their initial mass via winds at a flat rate from $z=4.5$ to 
the present.  

The self--similar form for CDM cluster density profiles seen in earlier
works (NFW2; Metzler 1995; Cole \& Lacey 1996;
Tormen, Bouchet \& White 1997) provides an excellent description of the
dark matter distribution in these simulated clusters.  The degree of
central concentration increases with decreasing cluster mass, as
expected from these earlier works and from the formation history of
objects in hierarchical clustering models.  The characteristic scale radius
for this profile appears at $0.1$--$0.2$ $r_{170}$ in these simulations;
however, Navarro, Frenk \& White (1997) caution that the location of the scale
radius is dependent upon the assumed value of the density parameter.

The two--fluid models, lacking winds, exhibit a self--similar gas
distribution.  At large radii, the gas density profile matches well the
expection from hydrostatic equilibrium and the self--similar profile
observed for the dark matter.  The agreement at smaller radii is
difficult to determine because of resolution limits.  The
introduction of winds raises the gas entropy above levels achieved by 
gravitational infall and thus produces a more extended gas
distribution within the dark matter dominated potential.  
The effect of winds is strongest on
low--mass clusters,
where the energy input by winds is comparable to the total thermal
energy of the gas.  Thus, the self--similarity seen in the gas
distribution of the two--fluid models is broken with the introduction
of winds, with the gas in low temperature clusters being more strongly
affected than their high temperature counterparts.

The energy input through winds is primarily spent in redistributing
the gas within the potential; the effect on the gas temperatures
is slight.  Thus, the introduction of winds does not seem to
affect the relation between ICM temperature and mass.  This, in
turn, supports the use of mass estimators that assume gas temperature
to be simply related to potential well depth.  However, the
dependence of this result upon the wind model chosen remains to
be determined.  Detailed temperature information from X--ray
satellites, combined with independent mass estimates from weak
lensing analysis, can probe this relation in real clusters.

Galaxies are more centrally concentrated than the dark matter in these
simulations.  Part of this effect arises simply because the initial
distribution of galaxies is more concentrated than the dark matter ---
an artifact of considering overdense peaks as likely sites for galaxy
formation.  However, in addition, a persistent velocity
bias between galaxies and dark matter is present in the ejection
ensemble; the degree of bias correlates with cluster mass in a
manner consistent with the expectations of bias induced by dynamical
friction.

As the gas is more extended than the dark matter, while the galaxies are
more concentrated, a gradient between galaxies and gas exists.  As a
result, metal abundance gradients are generic to the ejection models.
However, this result is sensitive to the specific wind model used;
the gradient is reduced if winds and metal enrichment occur only at
high redshift.  The strength of the metal abundance gradients seen in 
this work
shows no dependence on cluster mass, which may be in contradiction with
observations.  The relative ordering of extents of the three fluids
simulated is, however, consistent with present observations.

While energy input from feedback depresses central gas densities and causes
a more extended gas distribution, the effect is not significant
enough to explain observed cluster gas density cores observed in X--ray
images.  While cores of appropriate size are present in these simulations,
they appear predominantly numerical in origin.  Possible sources of
X--ray cores include additional sources of support such as magnetic
fields, or a change to the cluster density profile such as
is expected for low--$\Omega$ cosmologies (Navarro, Frenk \& White 1997).
An ejection history featuring vigorous, early winds may also lead to
larger cores, since the central gas entropy could be raised above
that seen in the experiments presented here.  

Since the low--temperature clusters simulated experience the strongest
effect upon the gas density profile, estimates of the baryon fraction
should be taken from high--temperature clusters whenever possible.
The local baryon fraction in 10 keV clusters is approximately 90\%
of the global value and is insensitive to winds, while loss
approaching a factor of two in gas fraction can occur in interior parts
of low--temperature clusters.

This research suggests at least two directions for future numerical 
experiments.  One is to 
consider the simplified evolution of gas assumed to be isentropic 
at some high redshift, but which is allowed to change its adiabat 
through shock heating and (optionally) radiative cooling at later
times.  Systematically varying the initial adiabat would mimic the
effect of abrupt wind input of varying strength at high redshift, and
would allow structural issues, such as core radii generation, 
to be addressed.  
Another approach is to increase the spatial and mass resolution in the
experiments and add appropriate physics to allow galaxy formation 
to be modelled self--consistently within forming clusters.  This is 
the long--term goal of such cosmological simulations, but it remains 
a formidable task because of the uncertainties in modeling star
formation on galactic scales, the inherent complexity of the dynamical
system involved, and the large parameter space (physical and
numerical) associated with the problem.  There is much yet to be
gained from simple models, but the problem cannot be considered
``solved'' until the latter approach is complete.

\acknowledgments

CAM would like to thank Tina Bird, Mary Crone, Richard Mushotzky,
Julio Navarro, and Gordon Squires for useful discussions.
This work was supported by the NASA Astrophysics Theory Program
through grants NAGW--2367 and NAG5-2790.  CAM
also would like to acknowledge support from a Rackham Predoctoral
Fellowship and a Sigma Xi Grant--in--Aid of Research at the
University of Michigan, and the Department of Energy and NASA
Grant NAG5-2788 at Fermilab.  AEE acknowledges support from the CIES 
and CNRS during a sabbatical stay at the Institut d'Astrophysique in
Paris.

\appendix

\section{Appendix}

We provide here a brief summary of the hydrostatic, isothermal 
$\beta$--model (Cavaliere \& Fusco--Femiano 1976, 1978; Sarazin 
\& Bahcall, 1977).  Those familiar with the model may still 
wish to briefly review this appendix in order to become familiar with
the notation used in the paper.  

The model presumes that the intracluster medium is isothermal and 
hydrostatic in a potential well determined by collisionless, dark
matter following the density profile
\beq
\label{eq:betadmden}
\rho\,=\,
\rho_{0}\left[1\,+\,\left(\frac{r}{r_c}\right)^2\right]^{-3\alpha_{DM}/2},
\eeq
where $r_c$ is the core radius of the cluster.  All the early
literature on the model assumes that the collisionless matter 
is described by the King approximation to the isothermal sphere, meaning
$\alpha_{DM}\,=\,1$.  The relevant hydrostatic equations for the 
gas and collisionless matter are
\beq
\label{eq:gashydroeq}
\frac{dP}{dr}\,=\,-\rho_g\frac{d\Phi}{dr},
\eeq
and
\beq
\label{eq:dmhydroeq}
\frac{d\left(\rho_{DM} \sigma_r^2\right)}{dr}
\,+\,\frac{2\rho_{DM}}{r}\left[\sigma_r^2\,-\,\sigma_t^2\right]
\,=\,-\rho_{DM}\frac{d\Phi}{dr},
\eeq
with $P$ the gas thermal pressure; $\Phi$ the gravitational potential;
$\rho_g$ and $\rho_{DM}$ the densities of gas and collisionless
matter respectively; $\sigma_r$ the radial velocity dispersion, and
$\sigma_t\,=\,0.5\left(\sigma^2\,-\,\sigma_r^2\right)^{1/2}$ the
tangential velocity dispersion, of the collisionless matter.
The model assumes spherical symmetry, so that
$d\Phi/dr\,=\,-GM_{tot}\left(<r\right)/r^2$.  If one then demands
isothermality (that is, that the gas temperature and collisionless
matter velocity dispersion are independent of radius), one obtains
\beq
\label{eq:gashydroisoeq}
\frac{kT}{\mu m_p}\frac{d\rho_g}{dr}
\,=\,-\rho_g\frac{GM_{tot}\left(<r\right)}{r^2}
\eeq
and
\beq
\label{eq:dmhydroisoeq}
\sigma_r^2\frac{d\rho_{DM}}{dr}
\,+\,\frac{2\rho_{DM}\sigma_r^2}{r}A\left(r\right)
\,=\,-\rho_{DM}\frac{GM_{tot}\left(<r\right)}{r^2},
\eeq
where $A\left(r\right)\,=\,1\,-\,\sigma_t^2/\sigma_r^2$ is the 
anisotropy parameter.  If we assume that the collisionless matter follows
Equation~\ref{eq:betadmden}, then the gas density has a similar functional
form,
\beq
\label{eq:betagasden}
\rho_g\left(r\right)\,=\,
\rho_{g,0} \left[1 + \left(\frac{r}{r_c}\right)^2\right]^{-3\alpha_{GAS} / 2}.
\eeq
with 
\beq
\label{eq:alphagas}
\alpha_{GAS}\,=\,\beta_r\left(\alpha_{DM}\,-\,2/3 A\left(r\right)\right)
\eeq
and $\beta_r$ is the ratio of specific energies,
\beq
\label{eq:beta_rdef}
\beta_r\,=\,\frac{\sigma_r^2}{\left(\frac{kT}{\mu m_p}\right)}.
\eeq
All of these parameters can depend on radius in the cluster.  In the
standard implementation of the $\beta$--model, one assumes  
$\alpha_{DM}\,=\,1$  and an isotropic velocity dispersion $\sigma_r =
\sigma_t \equiv \sigma$, so 
$A\left(r\right)\,=\,0$ .  In this case, $\alpha_{GAS}\,=\,\beta$ of 
Equation~\ref{eq:betadef}.

The volume emissivity of
thermal bremsstrahlung emission is $\epsilon_{ff}\,\propto\,\rho_g^2T^{1/2}$;
the surface brightness is then calculated from a line--of--sight integral
of the square of the gas density.  The functional form used by observers
to fit X--ray surface brightness profiles is
\beq
\label{eq:betasbprof}
\Sigma_{x}\left(\theta\right) \,=\, \Sigma_{0}
        \left[ 1\,+\,\left(\theta / \theta_{x}\right)^{2}\right]
        ^{-3\beta_{fit}\,+\,1/2},
\eeq
which is identical to the result obtained from the $\beta$--model
profile if the cluster is isothermal, with $\beta_{fit}\,=\,\alpha_{GAS}$,
and $\theta_x$ being the angular size of the core radius $r_c$ at the
cluster's redshift.  Typically, data show $\beta_{fit}\,<\,\beta$, where
the ratio of specific energies $\beta$ is determined by direct
spectral analysis of cluster galaxy redshifts and by X--ray spectral
fitting.  Various effects have been noted to explain this so--called
``$\beta$--discrepancy''.  Since the logarithmic slope of
Equation~(\ref{eq:betasbprof}) is a function of radius,
$\beta_{fit}\,<\,\alpha_{GAS}$ unless the data used in the surface
brightness fit extends to sufficient radius.  Also,
from the relation between $\alpha_{GAS}$ and $\beta$,
$\alpha_{GAS}$ will be less than $\beta$ if $\alpha_{DM}\,<\,1$ or if
$A\left(r\right)\,>\,0$, both of which are seen in simulations
(Evrard 1990a,b).  Finally, if thermalization of cluster gas after
infall is incomplete, and the ICM is partially supported by residual
motions, then $\beta$ overestimates the ratio of specific energies,
and the ``effective'' value of $\beta$ is decreased.

The galaxies, as another
collisionless fluid, should also follow this functional form, although
perhaps with a different large--radius logarithmic slope to the density
profile --- that is, $\alpha_{gal}$ need not equal $\alpha_{DM}$.

\pagebreak
\vfill\eject

\clearpage

\begin{figure}
\caption{\label{fig:dm_vrprof} Dark matter radial velocity profiles at
$z\,=\,0.02$ for
four members of the two--fluid ensemble.  Profiles are centered on the most
bound dark matter particle of each cluster.  Mean interior density contrast
is used as the radial coordinate.  The dashed lines
mark the predictions of the simple spherical infall model for the virial
radius and the turnaround radius.}
\end{figure}

\begin{figure}
\caption{\label{fig:dm_denprof} Dark matter density profiles for the
eighteen members of the two--fluid ensemble,
at $z\,=\,0.02$.  The vertical lines mark the value of the gravitational
softening parameter $\epsilon$ for the various runs.}
\end{figure}

\begin{figure}
\caption{\label{fig:dm_denprof_sc}
(a) Rescaled dark matter density profiles for the two--fluid
ensemble at $z\,=\,0.02$.  Densities are scaled to
multiples of the background (critical) density, and radii
rescaled by the radius with a mean interior density contrast of 170,
for each cluster.  Vertical lines mark the gravitational softening for
the various
runs.  (b) The mean dark matter profile for the two--fluid ensemble, along 
with three fits.  The mean density profile, denoted by the points and
error bars, is derived from the average value of the density contrast at
that scaled radius amongst the members of the ensemble; the errorbars
come from the scatter in this value
amongst the ensemble.  The solid line is a fit to the one--parameter density
profile suggested by NFW2.  The dotted line is
a simple power--law fit.  The dashed line is a fit to the density profile
of the standard $\beta$--model form.}
\end{figure}

\begin{figure}
\caption{\label{fig:dm_sigprof_sc}
Dark matter velocity dispersion and orbital anisotropy
profiles for the two--fluid models at $z\,=\,0.02$.  The top half shows
dark matter velocity dispersion profiles for the eighteen two--fluid
runs.  The velocity dispersions have been rescaled by
$\sigma_{170}\,\simeq\,GM_{170}/2r_{170}$.}
The bottom half shows the value of the anisotropy parameter,
$A\left(r\right)\,=\,1\,-\,\sigma^2_t/\sigma^2_r$, for the runs.  The
radial coordinate has been rescaled by $r_{170}$.  
\end{figure}

\begin{figure}
\caption{\label{fig:galnumdenprofs}
Galaxy number density profiles for the six most massive clusters in the
ejection ensemble, at $z\,=\,0.02$.  Each radial bin holds five galaxies.
The solid lines show fits to the $\beta$--model profile; the values for
$3\alpha_{gal}$ shown give the large--radius logarithmic slope of the
number density profile from the fits.
}
\end{figure}

\begin{figure}
\caption{\label{fig:velbias_evol2}
Evolution of the velocity bias parameter $b_v\,=\,\sigma_{gal}/\sigma_{DM}$,
averaged
over the entire ejection ensemble and over high-- and low--mass subsets.
Time is written in terms of the endpoint of the simulations at $z\,=\,0$.}
\end{figure}

\begin{figure}
\caption{\label{fig:gas_vrprof}
Gas radial mach number at $z\,=\,0.02$ for the eighteen members
of the two--fluid ensemble.  }
\end{figure}

\begin{figure}
\caption{\label{fig:gas_denprof_sc_new}
(a) Scaled two--fluid ensemble gas density profiles at $z\,=\,0.02$.  Radii
are rescaled by $r_{170}$ and densities by the background  baryon density,
similar to the dark matter in Figure~\protect\ref{fig:dm_denprof_sc}.
The vertical lines
here denote the values of the central SPH smoothing length for the members
of the ensemble.  A mean profile, defined as for the dark matter by
taking an average amongst the ensemble members in Eulerian bins,
is shown as a heavy solid line.  (b) The equivalent profiles for the
ejection ensemble.  The heavy solid line again marks a mean profile for
the ensemble.  The heavy dashed line shows the mean profile for the two--fluid
ensemble from panel(a), to facilitate comparison.}
\end{figure}

\begin{figure}
\caption{\label{fig:alpha_T}
The gas outer slope parameter $\alpha_{gas}$ from fits to the 
standard form, Equation~\protect\ref{eq:betaden}, as a function of 
cluster temperature.  Open circles are models without ejection (2F
ensemble), while asterisks show models with winds (EJ ensemble).  
The similarity in this variable evident in the 2F ensemble is 
broken by the introduction of winds.}
\end{figure}

\begin{figure}
\caption{\label{fig:T_M170}
Mass weighted cluster temperature $T$ against mass $M_{170}$, both 
measured within the virial radius $r_{170}$.  Symbol types are the
same as in Figure~\protect\ref{fig:alpha_T}.  The solid line gives a fit to
the 2F data $T_{2f}(M) = 4.0 (M/10^{15} \msol)^{2/3} \kev$, while the
dahsed line is the mass--temperature relation expected in the ejection
models if the input wind energy is retained as thermal energy
(Equation~\protect\ref{Tej_T2f}).  }
\end{figure}

\begin{figure}
\caption{\label{fig:work_Mvir2f_paper}
(a) Estimates of the work required to lift the gas in each cluster from
its final configuration in the 2F realization to that in the EJ
realization plotted against galaxy mass within the virial radius.  The
solid line gives the energy input by those galaxies over the course of
the simulation.  See text for a discussion.  
(b) Ratio of the work estimate to the total gas thermal energy as a
function of total cluster mass.  The work performed by winds is 
comparable to the total thermal energy in low mass, low temperature
clusters.  }
\end{figure}

\begin{figure}
\caption{\label{fig:gas_Tprof_sc_new}
Scaled two--fluid ensemble temperature profiles at
$z\,=\,0.02$.  Here temperatures are scaled
by $T_{170}\,=\,\left(\mu m_p/k\right)\left(GM_{170}/2r_{170}\right)$.
The left--hand
panel shows the results for the two--fluid ensemble, while the right--hand
panel shows the results for the ejection ensemble.}
\end{figure}

\begin{figure}
\caption{\label{fig:betatruehists}
Histograms of the distribution of
$\beta_{DM}\,=\,\mu m_p\sigma_{DM}^2/kT$ for the two--fluid and ejection
ensembles, both with and without correcting temperatures to account
for gas specific energy in bulk motions, at $z\,=\,0.02$.}
\end{figure}

\begin{figure}
\caption{\label{fig:Feprof}
Iron abundance profiles for five members of the ejection
ensemble at $z\,=\,0.02$.  Also shown are mean profiles for the entire
ensemble (heavy solid line), the high mass subset (heavy dotted line), and
the low mass subset (heavy dashed line).}
\end{figure}

\begin{figure}
\caption{\label{fig:gasdenprofs5}
(a) $z\,=\,0.02$ mean gas density profile for the two--fluid
ensemble.  Error bars come from the dispersion in each Eulerian bin
amongst members of the ensemble.  The solid line is a fit to the form
predicted by the NFW model with the assumptions of hydrostatic equilibrium
and isothermality.  The dashed line is the prediction of the
NFW model for the scale length of the potential ($\lambda\,=\,0.154$) found
through studying the mean dark matter distribution, without including gas
motions in the pressure support.  The dash--dotted line is the prediction
of the generalized NFW model for the known form of the potential
{\it including} gas motions in the pressure support term.  These last
two lines are normalized to produce the correct amount of gas within
$r_{170}$.  (b) The equivalent plot for the ejection ensemble.  Again,
the solid line is a direct fit to the mean profile; the dashed line is
the prediction of the NFW model assuming the scale length found in the
potential; and the dash--dotted line is the prediction of the model when
including bulk motions in the temperature.  Finally, the direct fit
to the two--fluid model points is shown here as a dotted line, to
facilitate comparison.}
\end{figure}

\begin{figure}
\caption{\label{fig:gasden_corehists}
Histograms of gas density core radii at $z\,=\,0.02$, for both ensembles.
Values of the
core radius $r_{c}$ are taken from fits to the $\beta$--model form.
Shown are absolute values of $r_{c}$, as well as values scaled in terms
of $r_{170}$, the gravitational softening, and the SPH smoothing length
at the cluster center.}
\end{figure}

\begin{figure}
\caption{\label{fig:relextents_new}
Enclosed mass profiles for the various cluster components
at $z\,=\,0.02$.  Each line corresponds to the fraction of the mass in that
component at $r_{170}$ that is enclosed within radius $r$.  The clusters in
each
ensemble were renormalized to their values of $r_{170}$ and ``stacked'' on
top of each other to produce a characteristic curve for each component in
the ensemble.  Components are marked by different lines:  two--fluid ensemble
dark matter (solid); two--fluid ensemble gas (dotted); ejection ensemble
dark matter (short--dashed); ejection ensemble gas (long--dashed); and ejection
ensemble galaxies (dot--dashed).}
\end{figure}

\begin{figure}
\caption{\label{fig:fb_T_paper}
Local baryon fractions, normalized to the cosmic ratio, within two 
different density contrasts displayed versus mass--weighted temperature.  
Open circles show the 2F ensemble results, while components within 
the EJ ensemble are shown individually (filled triangles : gas; 
filled circles : galaxies) and in total (asterisks).  Solid lines show
fits to the EJ total, Equation~\protect\ref{eq:fb_T_fit} at each density
contrast.  The short--dashed line in the $\delta_c \se 500$ panel shows 
$\delta_c \se 170$ result; the baryon fraction decreases toward smaller
radii or larger density contrasts.  The long--dashed line at unity
reflects a local baryon fraction equal to the cosmic mean value.  
All clusters are baryon deficient.  The EJ ensemble shows a
dependence of the local baryon fraction on temperature but, at high
temperatures, the total baryon content is similar to that in the
models without ejection.  
}
\end{figure}

\pagebreak

\clearpage
\begin{deluxetable}{lc}
\tablecolumns{2}
\tablewidth{0pt}
\tablecaption{General Simulation Parameters \label{tab:params}}
\tablehead{
\colhead{Parameter} &
\colhead{Value}
}
\startdata
Comoving Box Length $L$ ($\mpc$)		& 20, 25, 30, 40, 60 \nl
Total Mass in Box ($\msol$)			& 
$4.44\times 10^{15}\left(\frac{L}{40 \mpc}\right)^3$\nl
Number of Dark Matter Particles			& 32768	\nl
Mass per Dark Matter Particle ($\msol$)		& 
$1.22\times 10^{11}\left(\frac{L}{40 \mpc}\right)^3$\nl
Number of Gas Particles in Two--Fluid Runs	& 32768 \nl
Mass per Gas Particle ($\msol$)			& 
$1.35\times 10^{10}\left(\frac{L}{40 \mpc}\right)^3$\nl
Initial Redshift				& 9 \nl
Timestep (yr)					& $1.29\times 10^{7}$\nl
Initial Temperature of Gas			& $10^{4}\K$ \nl
Specific Energy of Wind ($L_{wind}/\dot{M}$)	& $1.37\times 10^{8}\K$ \nl
\enddata
\end{deluxetable}

\begin{deluxetable}{lccc}
\tablecolumns{4}
\tablewidth{0pt}
\tablecaption{Final Characteristics of Two--Fluid Models \label{tab:2fchars}}
\tablehead{
\colhead{Run} &
\colhead{$M_{170}$} &
\colhead{$\sigma_{DM,170}$} &
\colhead{$T_{170}$} \\
\colhead{} &
\colhead{($\msol$)} &
\colhead{($\kms$)} &
\colhead{(K)} \nl
}
\startdata
b20bn & $8.34\times 10^{13}$ & $365$ & $8.75\times 10^{6}$ \nl
b20cn & $1.29\times 10^{14}$ & $435$ & $1.15\times 10^{7}$ \nl
b20en & $8.36\times 10^{13}$ & $353$ & $7.81\times 10^{6}$ \nl
b20fn & $1.16\times 10^{14}$ & $462$ & $1.28\times 10^{7}$ \nl
b25an & $2.84\times 10^{14}$ & $597$ & $2.00\times 10^{7}$ \nl
b25bn & $2.79\times 10^{14}$ & $564$ & $2.08\times 10^{7}$ \nl
b25cn & $3.31\times 10^{14}$ & $576$ & $2.07\times 10^{7}$ \nl
b25dn & $2.60\times 10^{14}$ & $564$ & $2.05\times 10^{7}$ \nl
b30an & $4.59\times 10^{14}$ & $665$ & $2.84\times 10^{7}$ \nl
b30bn & $4.94\times 10^{14}$ & $721$ & $3.10\times 10^{7}$ \nl
b30cn & $4.66\times 10^{14}$ & $667$ & $2.69\times 10^{7}$ \nl
b30dn & $4.57\times 10^{14}$ & $632$ & $2.64\times 10^{7}$ \nl
b40an & $1.14\times 10^{15}$ & $881$ & $4.94\times 10^{7}$ \nl
b40bn & $9.20\times 10^{14}$ & $818$ & $4.48\times 10^{7}$ \nl
b40cn & $1.04\times 10^{15}$ & $942$ & $5.28\times 10^{7}$ \nl
b60bn & $2.61\times 10^{15}$ & $1120$ & $8.55\times 10^{7}$ \nl
b60cn & $3.92\times 10^{15}$ & $1420$ & $1.12\times 10^{8}$ \nl
b60dn & $3.59\times 10^{15}$ & $1310$ & $9.58\times 10^{7}$ \nl
\enddata
\end{deluxetable}

\begin{deluxetable}{lcccc}
\tablecolumns{5}
\tablewidth{0pt}
\tablecaption{Final Characteristics of Ejection Models \label{tab:ejchars}}
\tablehead{
\colhead{Run} &
\colhead{$f_{gas}$} &
\colhead{$N_{GAL}$} &
\colhead{$N_{GAL}\left(< r_{170}\right)$} &
\colhead{$T_{170}$ (K)} 
}
\startdata
b20b & $0.914$ & 24 & 8 & $1.02\times 10^{7}$ \nl
b20c & $0.879$ & 35 & 16 & $1.31\times 10^{7}$ \nl
b20e & $0.921$ & 26 & 11 & $9.29\times 10^{6}$ \nl
b20f & $0.913$ & 39 & 15 & $1.34\times 10^{7}$ \nl
b25a & $0.912$ & 41 & 31 & $2.22\times 10^{7}$ \nl
b25b & $0.887$ & 48 & 31 & $2.24\times 10^{7}$ \nl
b25c & $0.881$ & 56 & 36 & $2.49\times 10^{7}$ \nl
b25d & $0.895$ & 50 & 24 & $2.17\times 10^{7}$ \nl
b30a & $0.879$ & 79 & 46 & $3.04\times 10^{7}$ \nl
b30b & $0.904$ & 64 & 42 & $3.25\times 10^{7}$ \nl
b30c & $0.899$ & 78 & 39 & $3.03\times 10^{7}$ \nl
b30d & $0.886$ & 79 & 40 & $2.93\times 10^{7}$ \nl
b40a & $0.875$ & 128 & 60 & $5.47\times 10^{7}$ \nl
b40b & $0.878$ & 128 & 55 & $4.74\times 10^{7}$ \nl
b40c & $0.894$ & 115 & 53 & $5.43\times 10^{7}$ \nl
b60b & $0.842$ & 201 & 72 & $9.21\times 10^{7}$ \nl
b60c & $0.849$ & 195 & 103 & $1.19\times 10^{8}$ \nl
b60d & $0.851$ & 189 & 100 & $1.01\times 10^{8}$ \nl
\enddata
\end{deluxetable}

\begin{deluxetable}{lccc}
\tablecolumns{4}
\tablewidth{0pt}
\tablecaption{Dark Matter Density Profile Fits \label{tab:ej2dmfits}}
\tablehead{
\colhead{} &
\colhead{NFW} &
\colhead{$\beta$--model} &
\colhead{Power-Law Slope $\alpha$} \nl
\colhead{Ensemble} &
\colhead{$\lambda$} &
\colhead{$\alpha_{DM}$} &
\colhead{$100\le\rho/\rho_{back}\le 3\times10^{3}$}
}
\startdata
& $0.154\pm0.008$ & $0.826\pm0.018$ & $2.39\pm0.08$ \nl
All Runs & $\chi^{2}/\nu\,=\,0.53$ & $\chi^{2}/\nu\,=\,0.39$ & $\chi^{2}/\nu\,=\,0.24$ \nl
& $q\,=\,0.99$ & $q\,=\,0.99$ & $q\,=\,1.0$ \nl
\hline
& $0.176\pm0.01$ & $0.816\pm0.018$ & $2.35\pm0.07$ \nl
High Mass & $\chi^{2}/\nu\,=\,0.75$ & $\chi^{2}/\nu\,=\,0.60$ & $\chi^{2}/\nu\,=\,0.53$ \nl
Subset & $q\,=\,0.85$ & $q\,=\,0.98$ & $q\,=\,0.92$ \nl
\hline
& $0.145\pm0.005$ & $0.835\pm0.015$ & $2.36\pm0.10$ \nl
Low Mass & $\chi^{2}/\nu\,=\,1.24$ & $\chi^{2}/\nu\,=\,0.587$ & $\chi^{2}/\nu\,=\,0.13$ \nl
Subset & $q\,=\,0.12$ & $q\,=\,0.98$ & $q\,=\,1.0$ \nl
\hline
\enddata
\end{deluxetable}

\begin{deluxetable}{ccc}
\tablecolumns{3}
\tablewidth{0pt}
\tablecaption{Mean ICM $\alpha_{GAS}$ Values \label{tab:alphagas}}
\tablehead{
\colhead{Sample} &
\colhead{2F Ensemble} &
\colhead{EJ Ensemble}
}
\startdata
All  & $0.870 \pm 0.002$ & $0.704 \pm 0.008$ \nl
$T<4$ keV & $0.860 \pm 0.002$ & $0.663 \pm 0.007$ \nl
$T>4$ keV & $0.891 \pm 0.002$ & $0.785 \pm 0.001$ \nl
\enddata
\end{deluxetable}

\begin{deluxetable}{lc}
\tablecolumns{2}
\tablewidth{0pt}
\tablecaption{Outer Density Profile Slopes \label{tab:logslopes}}
\tablehead{
\colhead{Component} &
\colhead{Value}
}
\startdata
Dark Matter & $-2.39$ \nl
Gas in two--fluid runs & $-2.34$ \nl
Gas in ejection runs & $-1.75$ \nl
Galaxies & $-2.22$ to $-4.88$ \nl
\enddata
\end{deluxetable}
\clearpage

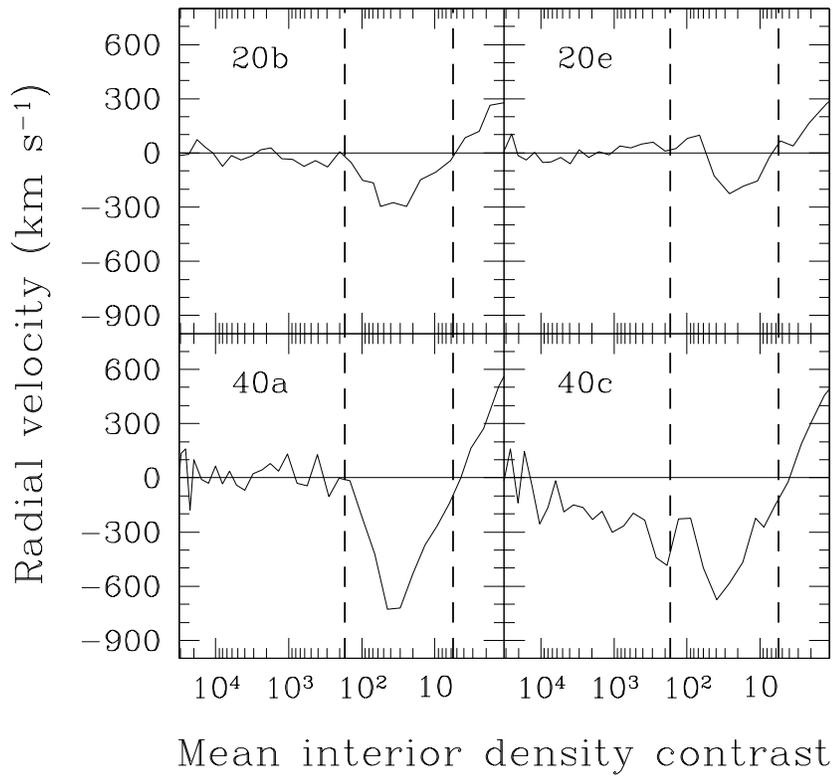
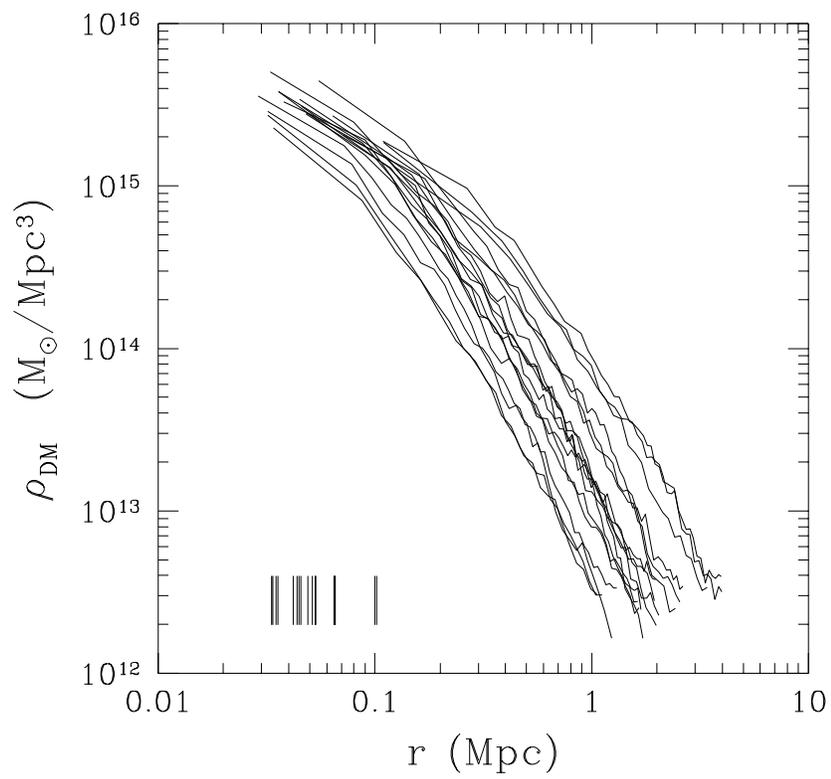
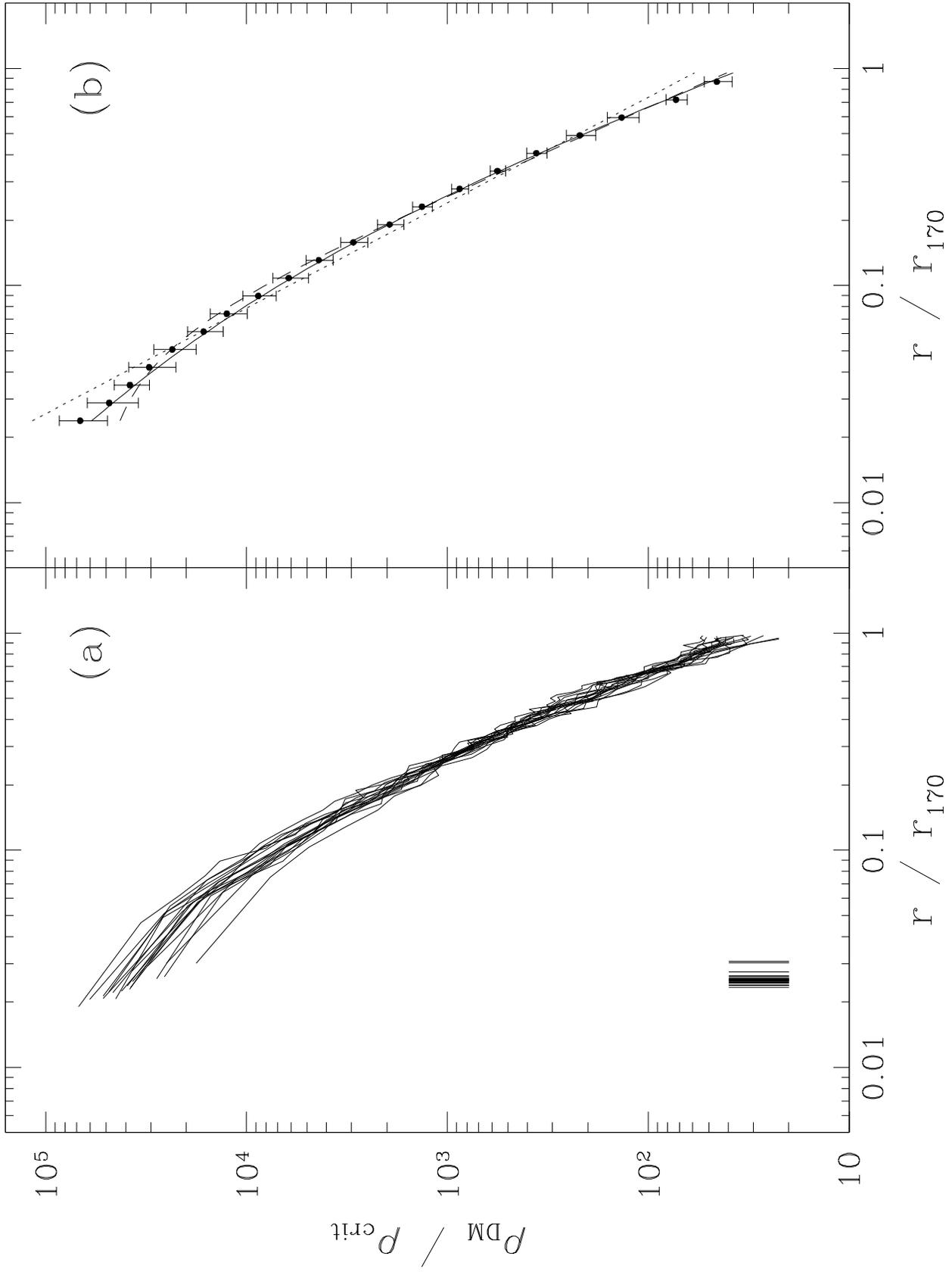
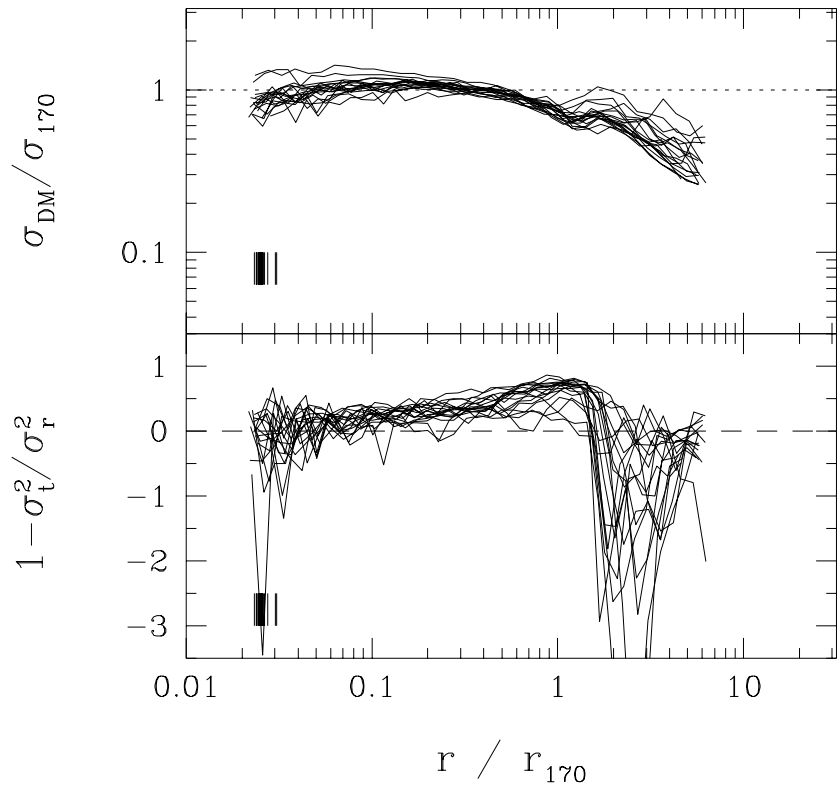
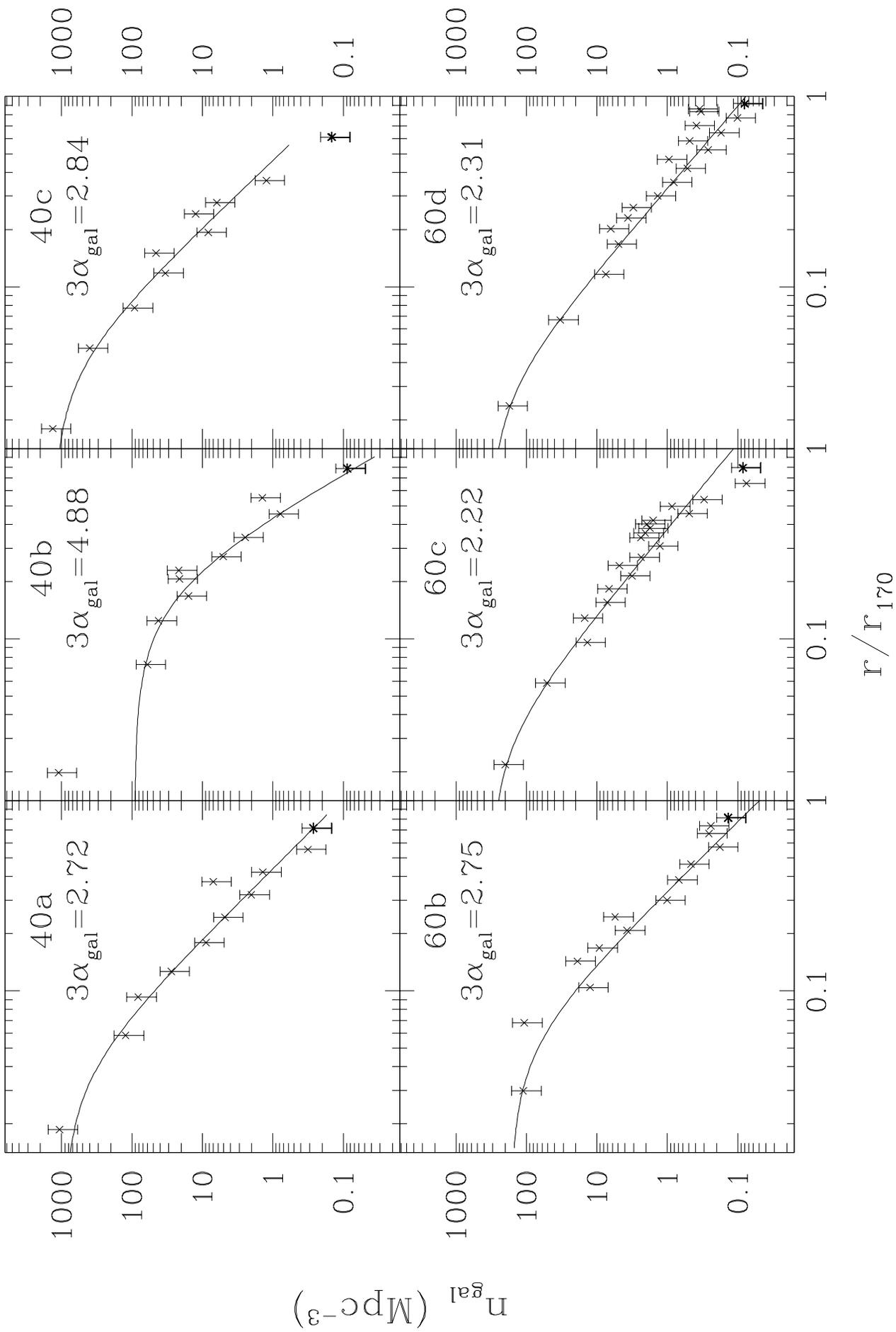
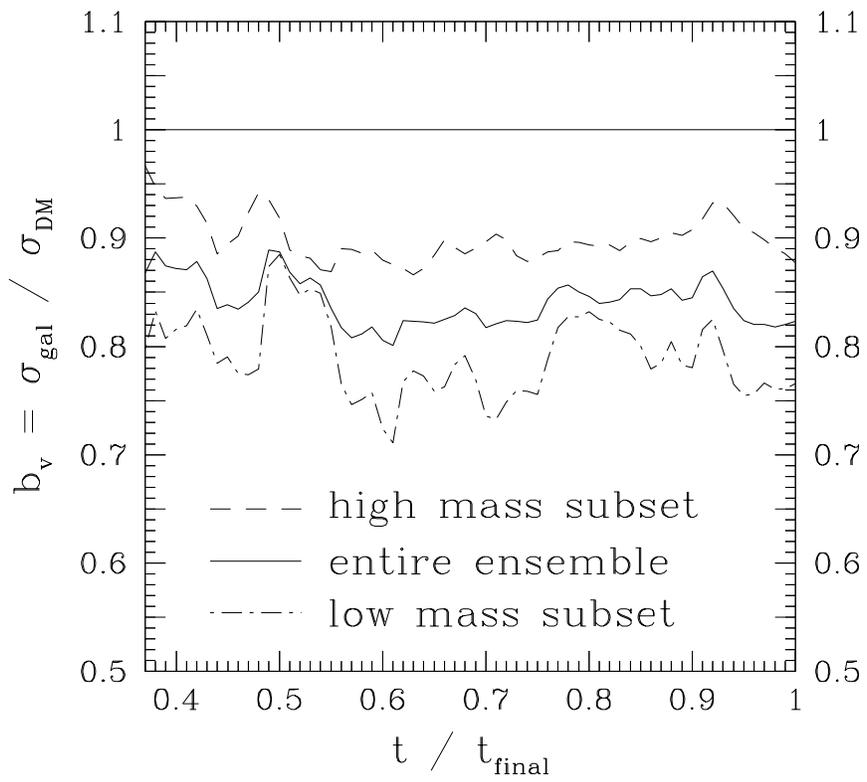
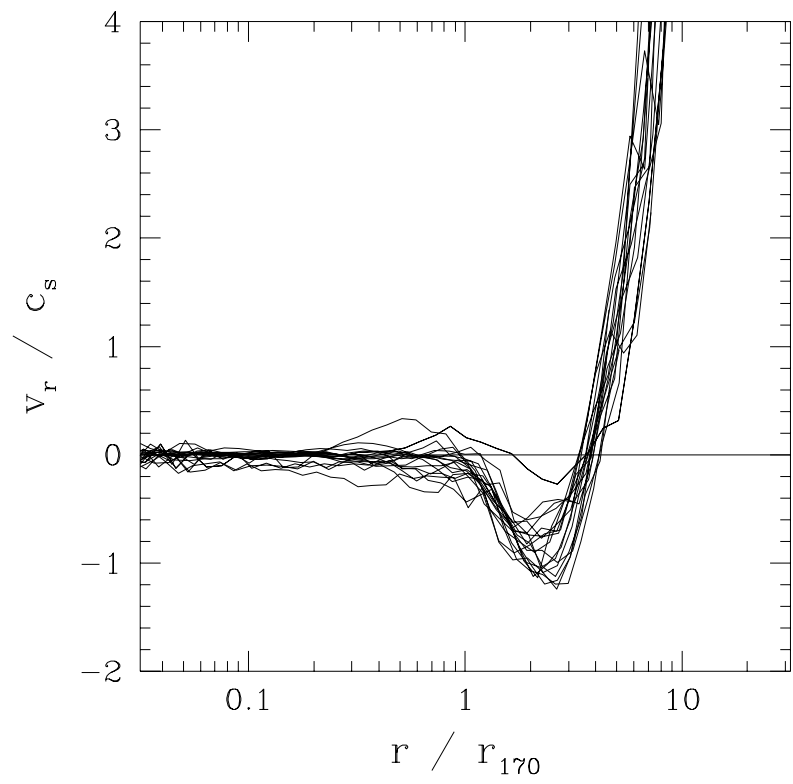
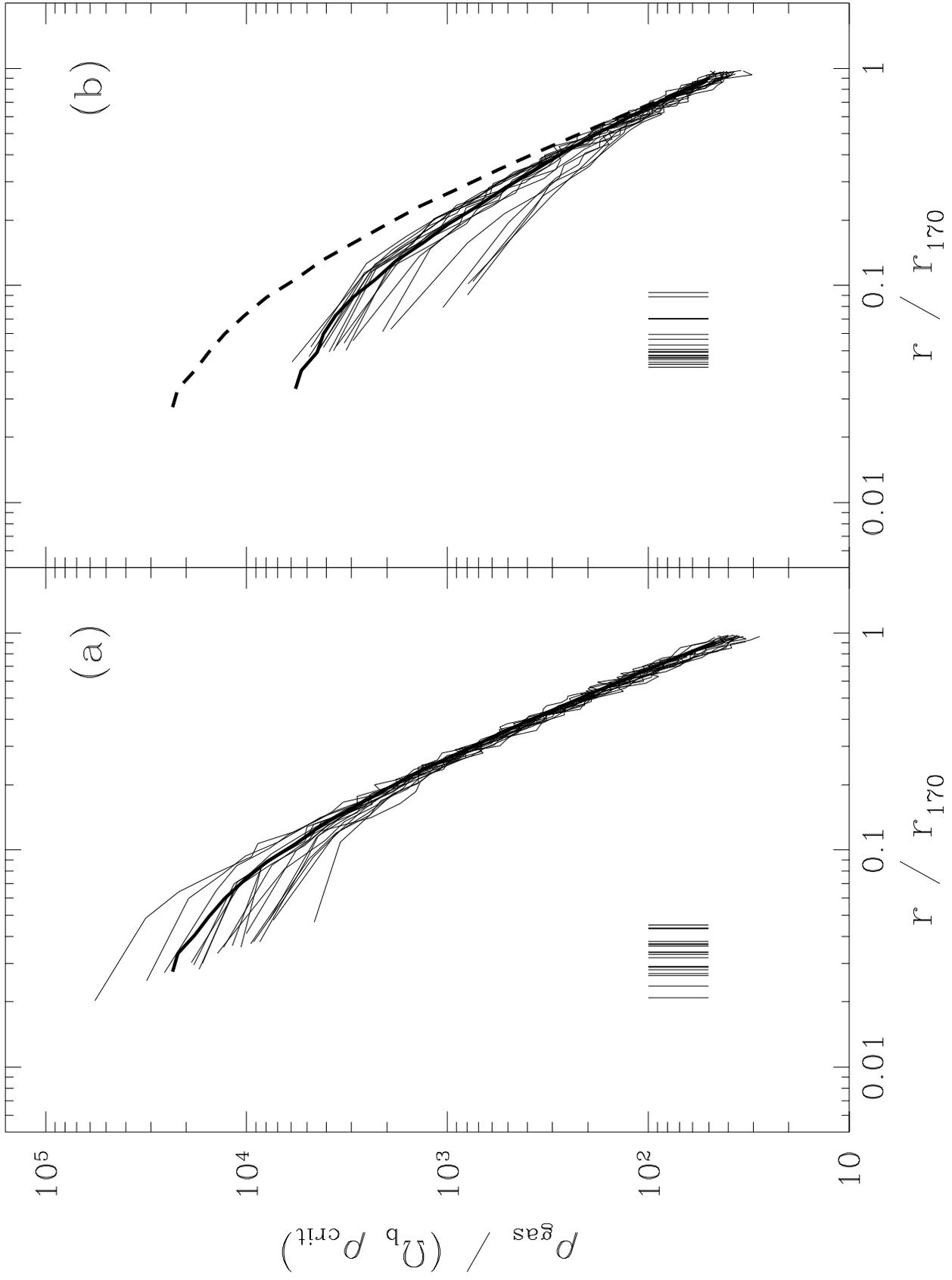
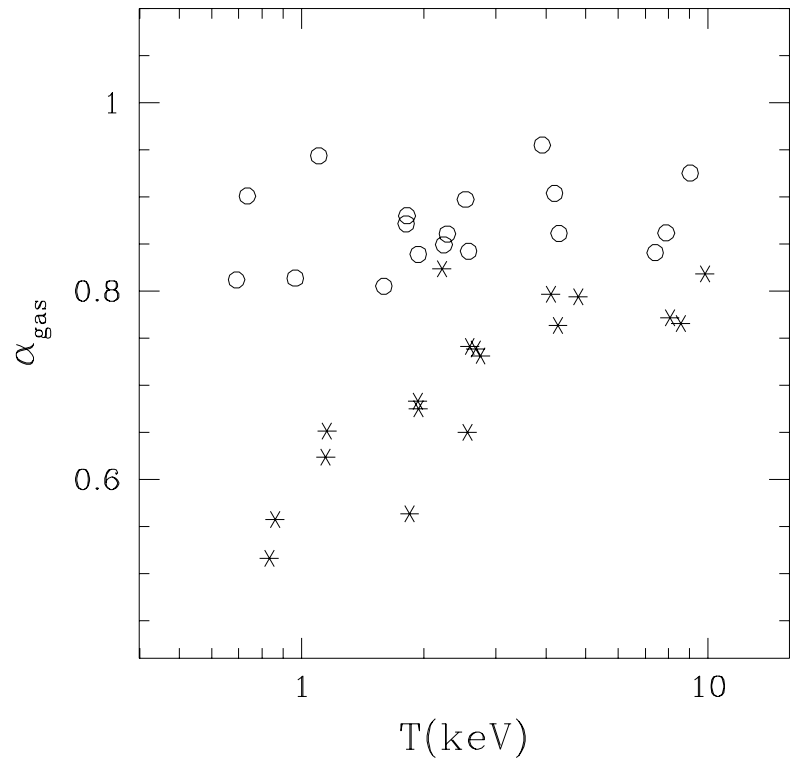
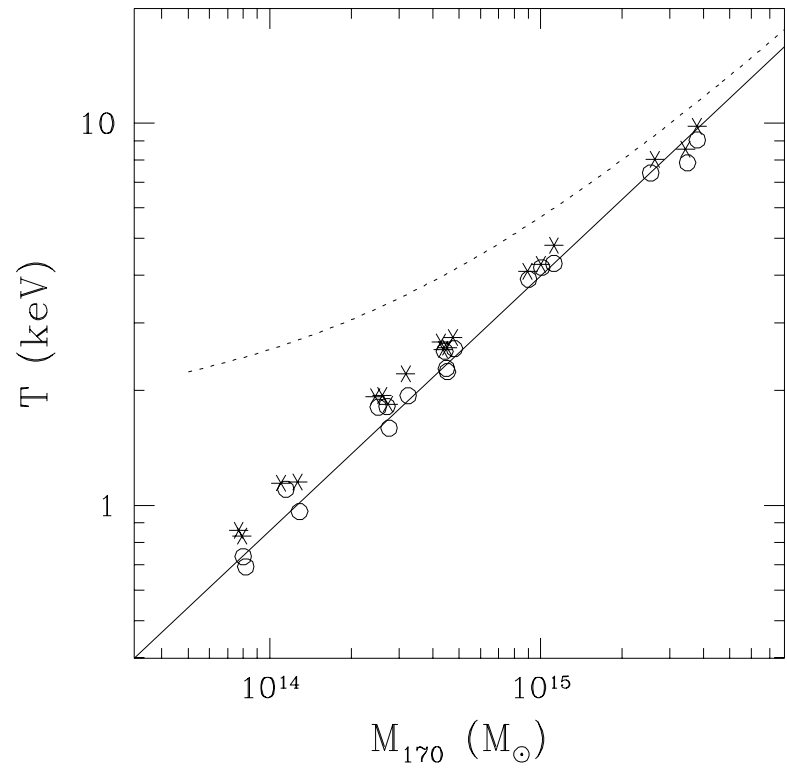
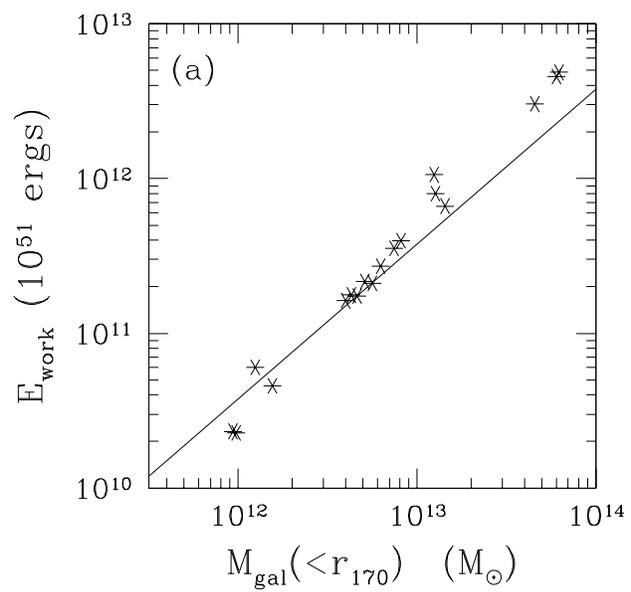
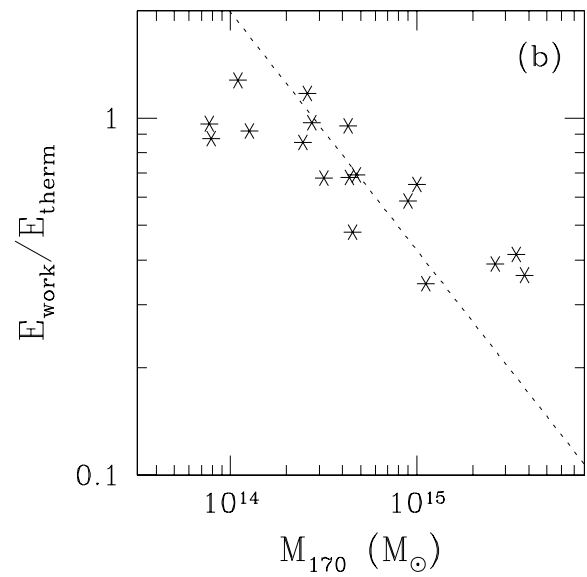
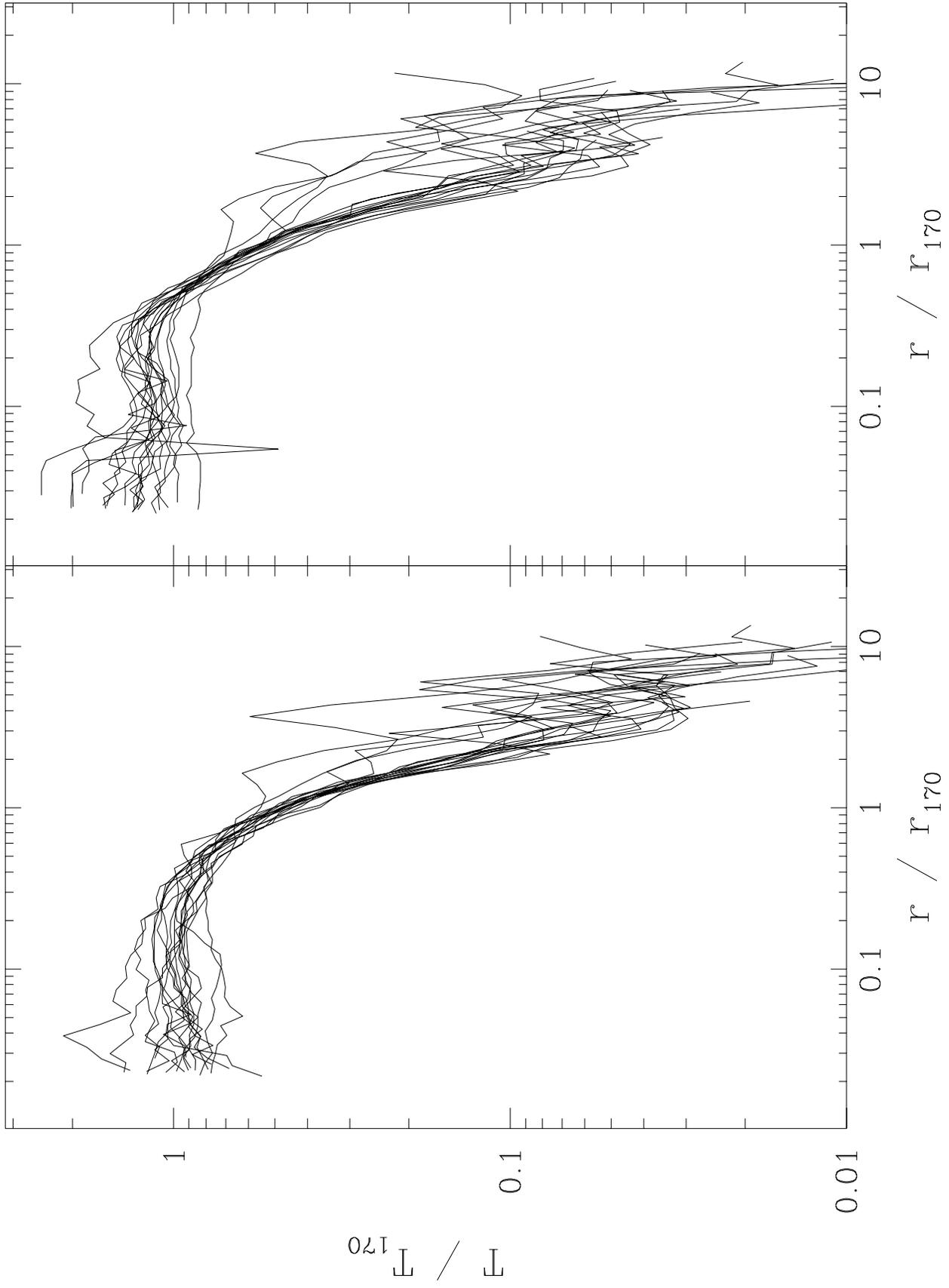
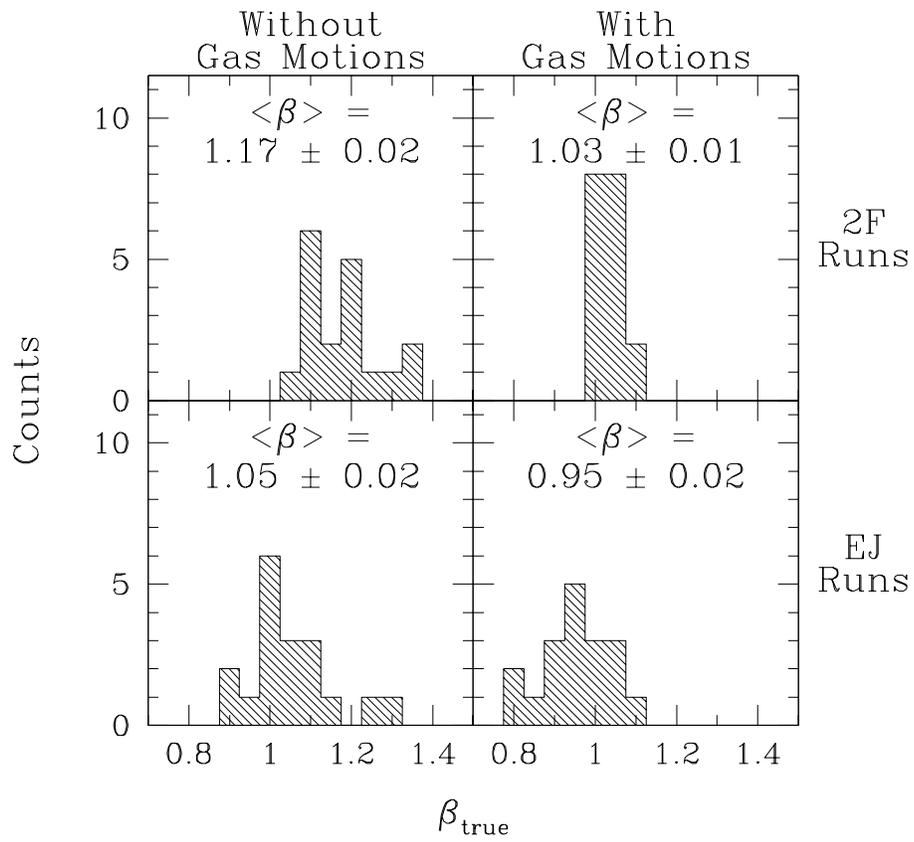
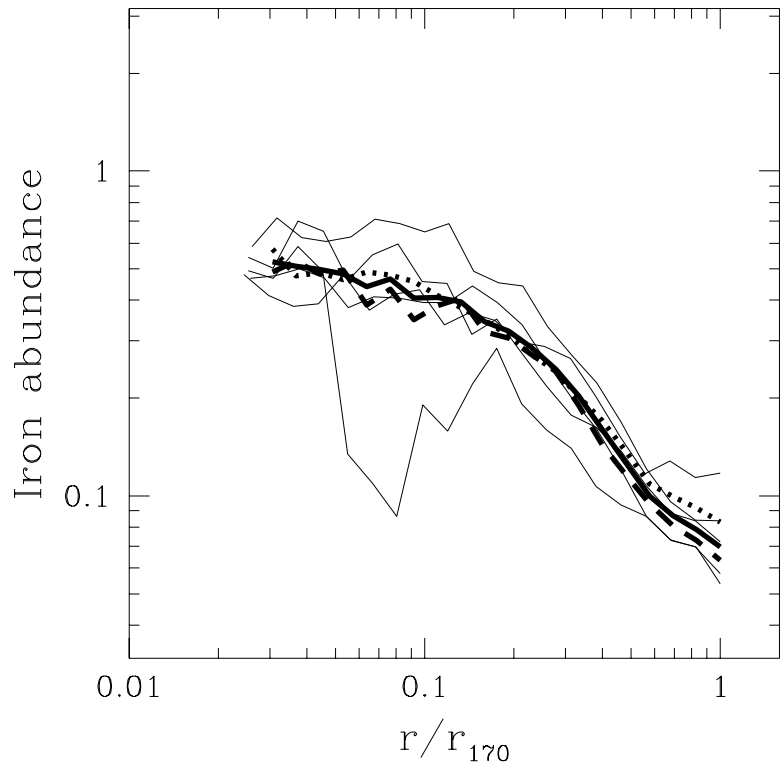
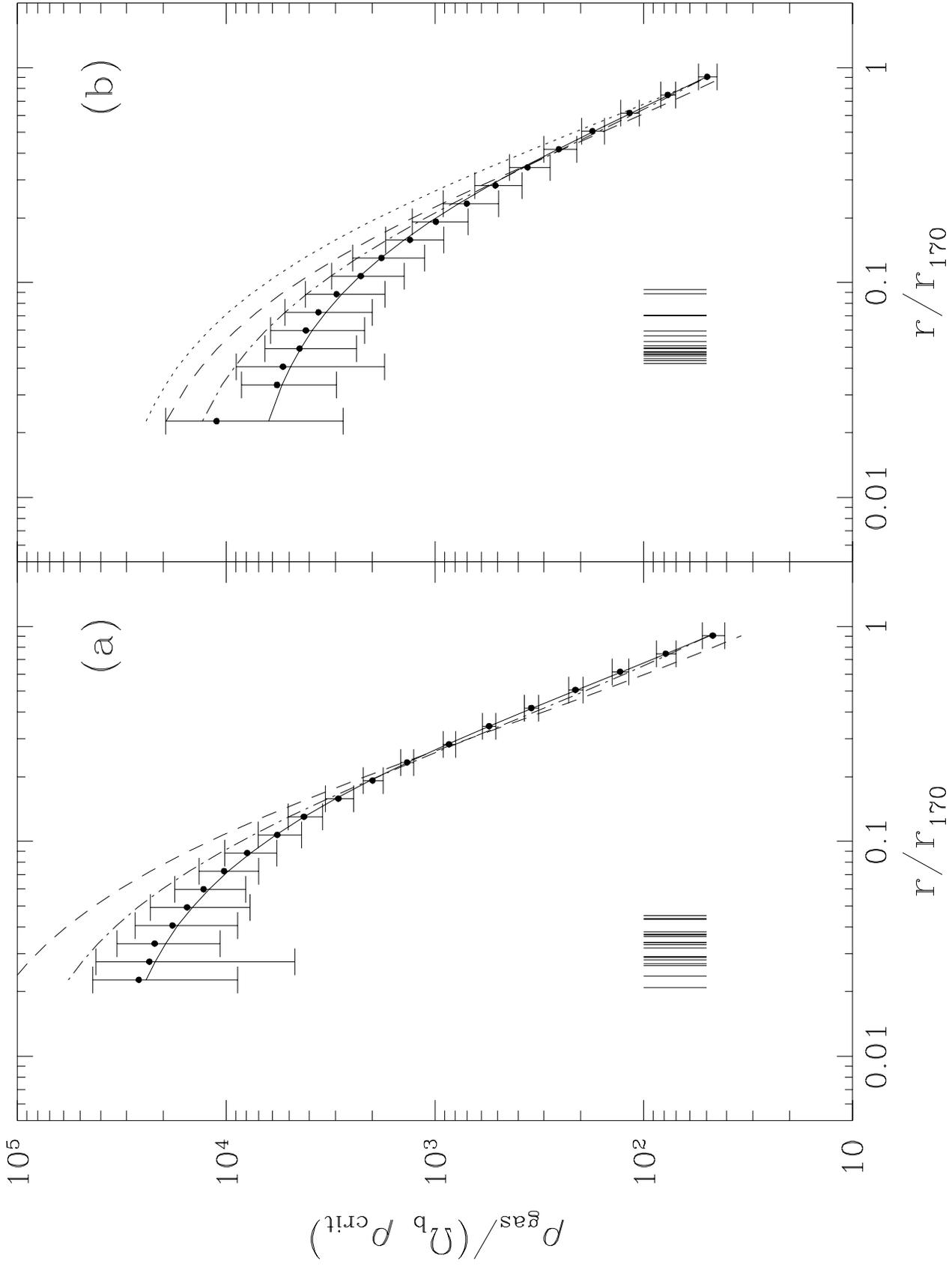
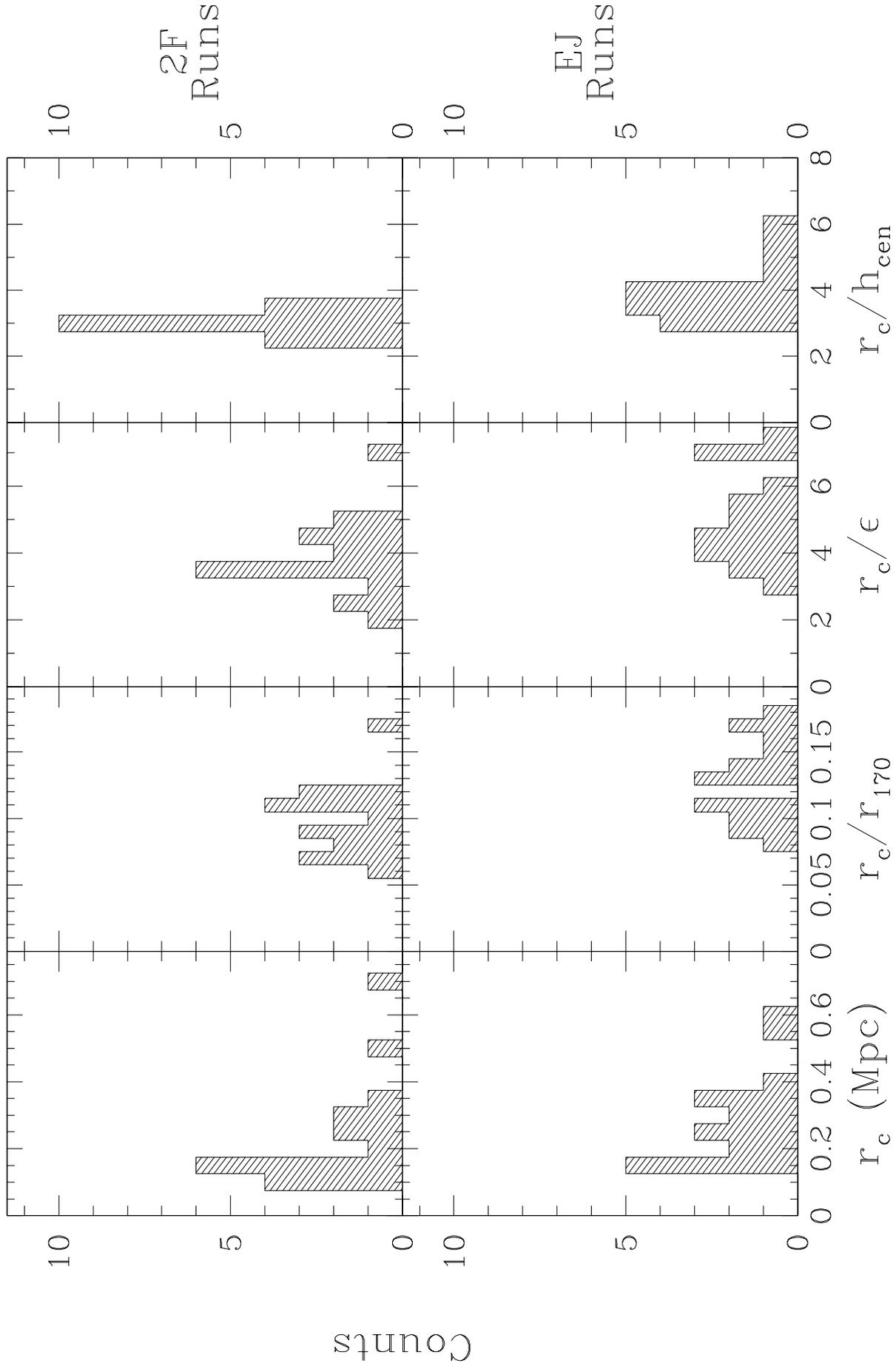
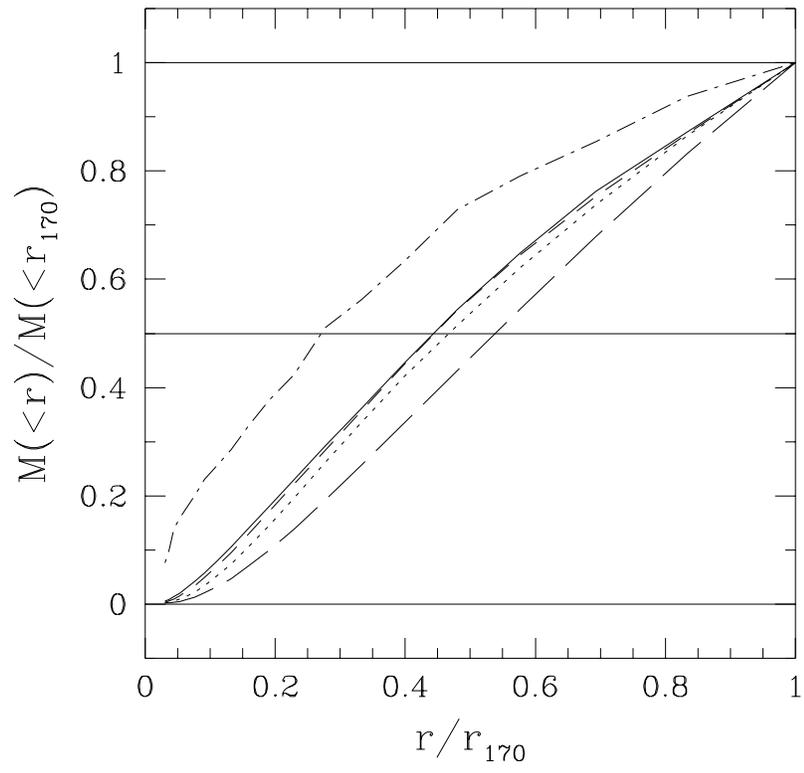
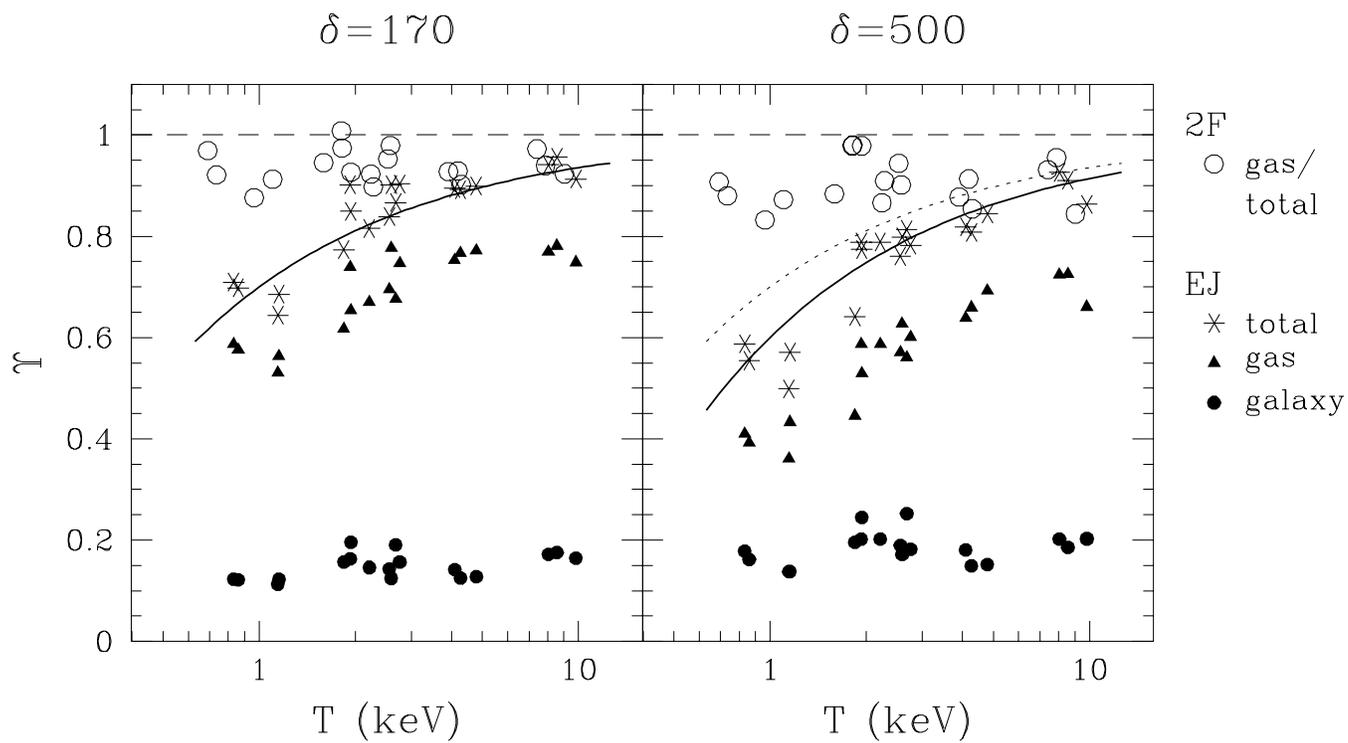
\end{document}